\documentclass[aps,prb,reprint,a4paper,floatfix,groupedaddress,amsmath,amssymb,amsfonts]{revtex4-1}
\usepackage{newtxtext,newtxmath}
\usepackage[utf8]{inputenx} 
\usepackage{graphicx}
\usepackage{xspace}
\usepackage[dvipsnames]{xcolor}
\usepackage[pdftex]{hyperref}
\hypersetup{
	pdftitle={Spin generation in completely MBE grown Co$_2$FeSi/MgO/GaAs lateral spin valves},
	colorlinks,
	citecolor=blue
	}
\usepackage[
,textwidth=17.5cm
,textheight=23.7cm
,verbose
,dvips
]{geometry}

\begin{document}
\title{Spin generation in completely MBE grown Co$_2$FeSi/MgO/GaAs lateral spin valves}

\author{G. Hoffmann}
\author{J. Herfort}
\author{M. Ramsteiner}
\email{ramsteiner@pdi-berlin.de}
\affiliation{Paul-Drude-Institut für Festkörperelektronik, Leibniz-Institut im Forschungsverbund Berlin e.V., Hausvogteiplatz 5-7, 10117 Berlin, Germany}

\begin{abstract}
We demonstrate first measurements of successful spin generation in crystalline Co$_2$FeSi/MgO/GaAs hybrid structures grown by molecular-beam epitaxy (MBE), with different MgO interlayer thicknesses. Using non-local spin valve and non-local Hanle measurement configurations, we determine spin lifetimes of ${\tau \approx 100}$~ns and spin diffusion lengths of ${\lambda \approx 5.6}$~$\mu$m for different MgO layer thicknesses proving the high quality of the GaAs transport channel. For an optimized MgO layer thickness, the bias dependence of the spin valve signals indicates the verification of the half-metallic gap (upper edge) of Co$_2$FeSi in accordance with first principle calculations. In addition to that, spin generation efficiencies up to 18$\%$ reveal the high potential of MgO interlayers at the Co$_2$FeSi/GaAs interface for further device applications.
\end{abstract}

\maketitle
\section{Introduction} \label{sec_intro}
The potential of semiconductor (SC) spin electronics lies in the control of both, charge and spin of electrons. The spin degree of freedom  allows to combine the conventional (charge related) data processing with non-volatile data storage in one device. \cite{Fabian2004,Datta1990a} In this context, the ferromagnet/semiconductor (FM/SC) hybrid system has been proven to be a very promising material system for future devices, \cite{Wurmehl2005, Manzke2013, Crooker2005a, Peterson2016} e.g. for a nonvolatile reconfigurable current divider.\cite{Manzke2013} The spin-polarized contact material Co$_2$FeSi (CFS) belongs to the group of Heusler alloys and is predicted to be halfmetallic, i.e., to be an ideal candidate for spin injection.\cite{Wurmehl2005, Webster1969} 
Regarding the FM/SC interface, the accurate design of the doping profile in the SC and the resulting electrical FM/SC contact characteristics are of major importance for efficient spin injection .\cite{Hu2011b, Manzke2017a}
However, the epitaxial growth of CFS on top of a SC suffers from Fe and Co atoms diffusing into the SC, which not only compensate the doping profile, but also form magnetic impurities by which the spin polarized electrons are scattered. The loss of spin information consequently leads to a deterioration of spintronic functionalities.\cite{Brandt2010,Hashimoto2007b, Jenichen2012a} The undesirable interdiffusion at the FM/SC interface can be eliminated by MgO interlayers as it has been shown, e.g., for the material system FM/MgO/GaAs.\cite{Hentschel2015a,Makarov2011b} Further advantages of MgO interlayers are their suitability as tunnel barriers, and also their usability as spin filters, in particular for non-halfmetallic FM contacts.\cite{Butler2005,Butler2008} Nonetheless, the spin generation in FM/MgO/SC tunnel contacts might be affected by FM and oxygen intermixing at the new FM/MgO interface when using non-epitaxially deposited MgO films \cite{Yang2011c}, as well as by oxygen vacancies in the MgO layer \cite{Guezo2008b} even during epitaxial growth.\cite{Yuasa2004}

In this paper, we discuss the impact of MgO interlayers on the electrical contact characteristics and spin generation in CFS/MgO/GaAs hybrid structures grown by molecular-beam epitaxy (MBE). The results obtained by (four terminal) non-local spin valve (NLSV) and non-local Hanle (NLH) measurements demonstrate the great advantage of MBE-grown MgO tunnel barriers in lateral CFS/MgO/GaAs spin valve (SV) systems compared to common FM/SC SV structures, while revealing some remaining challenges.
\section{Samples and Setup}
\begin{figure}[b!]
\centering
\includegraphics[width=8cm]{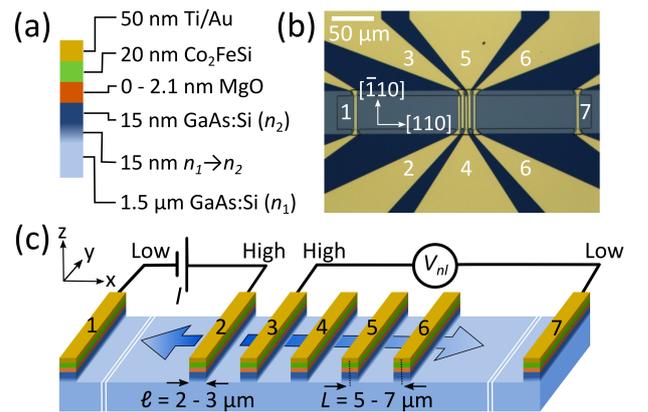}
\caption{(a) Relevant film sequence of the samples grown by MBE. (b)~Top view of a fabricated SV structure and orientation of the CFS strips with respect to the GaAs orientation. (c)~Schematic diagram of the CFS/MgO/GaAs structures for NLH and NLSV measurements. The blue arrows indicate the spin diffusion in the GaAs transport channel.}\label{fig1}
\end{figure}
For the growth of the CFS/MgO/GaAs hybrid structures a MBE system equipped with three chambers interconnected by ultra-high vacuum tubes was used for the separate growth of GaAs, MgO and CFS films. The samples were grown using semi-insulating GaAs(001) substrates. In Fig.~\ref{fig1}(a), the layer sequence of the relevant part is shown containing a 1.5-$\mu$m-thick lightly $n$-type GaAs:Si film with a Si concentration of ${n_1 = 5 \times 10^{16}}$~cm$^{-3}$, followed by a 15-nm thick $n$-type GaAs:Si film with gradually increasing Si doping (from ${n_1}$ to ${n_2= 5\times 10^{18}}$~cm$^{-3}$), and a 15-nm thick $n$-type GaAs:Si film with a Si concentration $n_2$. The MgO layer thickness was targeted to be ${\text{0~ML}\leq t_{\text{MgO}}\leq \text{10~ML}}$ in steps of 2~ML where 1~ML MgO corresponds to ${t_{\text{MgO}} = 0.21}$~nm. On top of the MgO layer a 20~nm thick CFS film was grown in the metal growth chamber at a substrate temperature of $T_{\text{sub}}=280^{\circ}$C (measured by a kSA BandiT system). A detailed insight into the growth process is given in Ref.~\onlinecite{Hoffmann2018a}, where we could demonstrate that the GaAs, the MgO as well as the CFS films are crystalline with preferential in-plane epitaxial relationship CFS(001)[110]~$\parallel$~MgO(001)[100]~$\parallel$~GaAs(001)[100]. For the electrical measurements, the samples underwent several etching and optical lithography processes which are described in Ref.~\onlinecite{Bruski2013c}. During the device preparation process of the lateral SV structures, a 50~nm thick Ti/Au film was sputtered onto the CFS film for the generation of the electric contacts.
The resulting device structures shown in Fig.~\ref{fig1}(b) comprises a ${50\times 400}$~$\mu$m$^2$ conductive mesa region (light blue) with Ti/Au strip contacts (yellow) [see Fig.~\ref{fig1}(a)]. A 100~nm thick SiO$_2$ film (dark blue) prevents leakage currents between the contacts. The CFS strips are orientated along the easy axis of their magnetization, which was determined to be parallel to the [\={1}10] direction of the underlying GaAs substrate using superconducting quantum interference device (SQUID) magnetometry.\cite{Hoffmann2018a} For the present experiments, we used the design shown in Fig.\ref{fig1}(c) with various center-to-center separations $L$ between the inner strips (No.~2-6) of five to seven~$\mu$m and respective strip widths $\ell$ of two to three $\mu$m (note that the strips No.~1 and 7 are approximately 150~$\mu$m away from the inner strips). The different center-to-center separations allow for the measurement of spin signals at different distances $L$ which are needed for a reliable determination of spin lifetimes and spin diffusion lengths. 
We performed the non-local measurements at $T$~=~20~K in a He exchange gas cryostat, using a Keithley 236 DC source unit for the spin generation (injection as well as extraction) in the GaAs channel, and a Keithley 2182 nanovoltmeter for spin detection. In this context, negative (positive) current values correspond to spin injection (extraction) as depicted in Fig.~\ref{fig1}(c).
\section{Results and Discussion}
The electrical characteristics of the lightly doped GaAs channel were obtained by Hall effect measurements. Independent of MgO film thickness, the charge carrier density $n$ and mobility $\mu$ of the GaAs channel at a temperature of 20~K were found to be ${1.5 \times 10^{-16}}$~cm$^{-3}$ and 4000~cm$^2$/Vs, respectively. Note that the chosen thickness of the GaAs transport channel is presumably too large to detect a possible influence of atomic interdiffusion during the MBE growth on the carrier density and mobility.

In
\begin{figure}[t!]
\centering
\includegraphics[width=7.5cm]{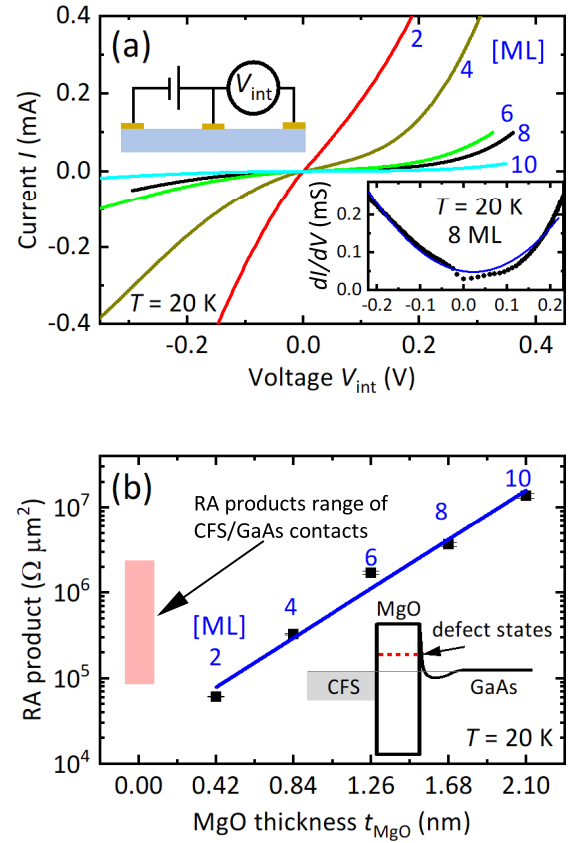}
\caption{(a) $I$-$V$ characteristics of samples with different MgO film thicknesses (2 to 10 ML) using the 3T arrangement depicted in the upper left inset. Lower right inset: Conductance $G=dI/dV$ as a function of $V_{\text{int}}$ for the 8~ML MgO sample. (b) RA product derived from $I$-$V$ curves near zero voltage as a function of MgO film thickness. Inset: reduced barrier height of the MgO tunnel barrier due to electrically active defect states. The red bar indicates the range of RA products from CFS/GaAs contacts produced under similar growth and procession conditions.\cite{Hoffmann2018a, Bruski2013c}}\label{fig2}
\end{figure} order to characterize the electrical behavior of individual CFS/MgO/GaAs contacts, a three-terminal (3T) arrangement was used [inset of Fig.~\ref{fig2}(a)] as described in Ref.~\onlinecite{Uemura2012c}. Fig.~\ref{fig2}(a) displays the resulting current-voltage ($I$-$V$) characteristics of SV structures comprising MgO films with thicknesses between 2 and 10~ML measured at $T$~=~20~K. The nearly symmetric shape of the $I$-$V$ curves is an indication that tunneling dominates the transport across the CFS/MgO/GaAs interfaces, which is an essential requirement for successful spin generation in the GaAs channel.\cite{Rashba2000, Fert2007} 
\begin{figure}[t]
\centering
\includegraphics[width=7cm]{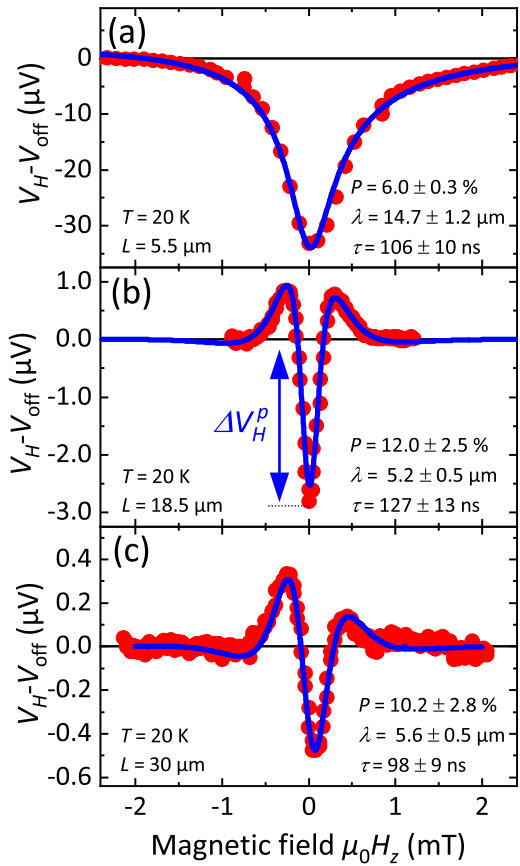}
\caption{NLH signal (red dots) of the sample with 4 ML MgO as a function of an external out-of-plane measured  at a bias current of -200~$\mu$A (spin injection condition) in the parallel magnetization configuration for center-to-center distances between injection and detection strips of (a)~5.5~$\mu$m, (b)~18.5~$\mu$m, and (c)~30~$\mu$m. For all Hanle curves, a linear background signal was subtracted. In addition, simulated Hanle curves resulting from fitting to Eq.~\ref{eq_hanle_fit_02} are shown (blue curves).}\label{fig3}
\end{figure}For small currents and in a linear transport regime, we determined contact resistance values and calculated the resistance-area (RA) products. The exponential dependence of the RA product as a function of the MgO layer thickness shown in Fig.~\ref{fig2}(b) is a clear signature for a tunnel contact, and corresponds to the first Rowell criterion: \cite{Hanbicki2003a,Jonsson-Akerman2002, Uemura2012c,Simmons1963,Butler2005}
\begin{align}
RA=R_0Ae^{-kt_{\text{MgO}}}.\label{eq_buttler}
\end{align}
Here, $R_0A$ is the RA product at ${t_{\text{MgO}}=0}$, and $k$ is the exponential decay constant. 
Our decay constant of ${k = 3.6}$~nm$^{-1}$ extracted from the data shown in Fig.~\ref{fig2}(b) is somewhat smaller than the one derived by Butler~et~al. who used first-principles based calculations for the Fe/MgO/Fe system, and who obtained a value of ${k = 5.6}$~nm$^{-1}$.\cite{Butler2005} The value derived by Butler~et~al. corresponds to a tunnel barrier height of only 0.3 to 0.4~eV in the framework of the Simmons approximation.\cite{Simmons1963} Similar barrier heights have also been found experimentally for various contacts with MgO tunnel barriers.\cite{Marukame2006, Yuasa2004, Uemura2012c} However, Butler~et~al. as well as Mavropoulos~et~al. pointed out that the Simmons model is not necessarily valid for the contact structures investigated here.\cite{Butler2005, Mavropoulos2000a} 
Our reduced tunneling decay constant can be understood in a qualitative manner by considering electrically active defects in the MgO films and near the CFS/MgO and MgO/GaAs interfaces, which reduce the effective barrier height as indicated by the inset of Fig.~\ref{fig2}(b). These defects  may be caused by oxygen vacancies \cite{Yuasa2004, Guezo2008b} and oxygen diffusion at the FM/MgO interface,\cite{Yang2011c} as well as Fe and Co diffusion across the CFS/MgO/GaAs interfaces. Furthermore, inhomogeneities in the MgO film thickness could be another reason for the reduced barrier height.\cite{Miller2006} Indeed, for our samples, we found evidence of small amounts of Fe and Co diffusion into the GaAs in the range of doping concentrations, and observed inhomogeneities in our MgO film thicknesses, in particular a waviness that occurred due to the interfacial strain caused by the lattice mismatch at the GaAs/MgO and MgO/CFS interfaces (for more details, see Ref.~\onlinecite{Hoffmann2018a}).

The second Rowell criterion for tunnel contacts requires a parabolic
voltage dependence of the conductance $G=dI/dV$.\cite{Brinkman1970} As demonstrated in the inset of Fig.~\ref{fig2}(a) for the 8~ML MgO sample, the experimental dependence of $G$ on $V_{\text{int}}$ 
can be well explained by  the function given in the framework of
the Brinkman-Dynes-Rowell model [blue line in the inset of Fig.~\ref{fig2}(a)] as commonly observed for various types
of tunnel contacts.\cite{Hanbicki2003a, Brinkman1970, Miao2009}
In general, $G(V_{\text{int}})$ is found to approach the ideal symmetric parabolic behavior with increasing MgO thickness (not shown here). The dip in $G(V_{\text{int}})$ at low $V_{\text{int}}$ is well-known as the so-called zero-bias anomaly. \cite{Oliver2004,Miao2009,Moodera1998,Yoo2010,Appelbaom1960} Since the zero-bias resistance of our CFS/MgO/GaAs contacts is found to exhibit a weak temperature dependence (increase of a factor 2 to 6.5 between 20 and 295~K), the third Rowell criterion for tunnel contacts is also fulfilled.\cite{Brinkman1970}

The shaded area in Fig.\ref{fig2}(b) indicates the range of RA products obtained for contacts without MgO interlayers fabricated under otherwise similar conditions.\cite{Hoffmann2018a, Bruski2013c} The relatively large scatter of RA products is attributed to the significant atomic diffusion of Co and Fe into the $n$-type doped GaAs during MBE growth. These diffusion processes and the corresponding electrical compensation of the $n$-type doping profile by Co and Fe impurities are strongly temperature dependent.\cite{Ramsteiner2008} Consequently, a subtle sample-to-sample variation of the actual growth temperature leads to sizeable changes in the RA products. In fact, the clear dependence of the RA product on the MgO film thickness shown in Fig.~\ref{fig2}(b) demonstrates the intended benefit of the MgO diffusion barrier regarding the adjustability of the contact characteristics.

The spin-related properties in the CFS/MgO/GaAs hybrid structures were investigated by NLH and NLSV measurements.\cite{Lou2007a, Uemura2011a} For this purpose, non-local voltages $V_{nl}$ [cf.~Fig.~\ref{fig1}(a)] were recorded as a function of the applied magnetic field (${V_{nl}=V_H}$ for NLH, ${V_{nl}=V_{SV}}$ for NLSV measurements). We performed Hanle measurements aligning the magnetization of spin generation and spin detection strip in parallel (p) and antiparallel (ap) configurations. In our measurements, we further observed a background signal linear in $H_y$ for NLSV, and parabolic in $H_z$ for NLH measurements in accordance with the observations of Ref.~\onlinecite{Ciorga2009,Lou2007a, Salis2009,Uemura2011a,Sasaki2010}. The origin of this background signal is not fully understood yet and is still under debate.\cite{Ciorga2009,Lou2007a,Ji2007} Due to the small fields used for the NLH measurements, we subtracted a linear background signal for both NLH and NLSV data. \begin{figure}[b!]
\centering
\includegraphics[width=7.5cm]{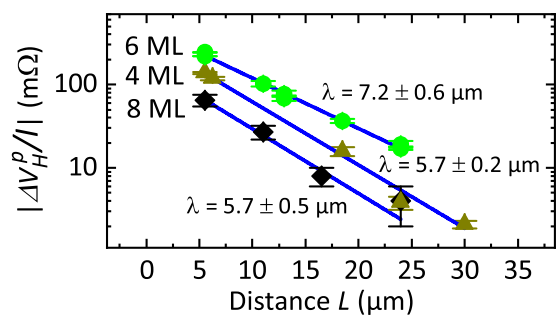}
\caption{Signal height $\vert{\Delta V_{H}^p/I}\vert$ of the Hanle curves in p strip configuration as a function of strip distance for different MgO layer thicknesses of 4,~6,~and~8~MLs. Spin diffusion lengths were derived by fitting the data using Eq.~\ref{eq_nlsv_distance_02} (blue lines).}\label{fig4}
\end{figure}
NLH curves for parallel strip configuration shown in Fig.~\ref{fig3}(a)-(c) provide evidence for successful spin generation and spin transport in the GaAs channel of the sample with MgO film thickness of 4~ML under spin-injection conditions. In Fig.~\ref{fig3}(b) and (c), oscillations in the Hanle signal are observable due to the precession of the electron spin around the axis of the applied magnetic field $H_z$.
The Hanle curves obtained for different separations of injector and detector strips can be well described by the commonly used expression:\cite{Fabian2007, Johnson1988}
\begin{align} 
\begin{split}
V_{H}(H_\bot)= &\frac{\rho}{S}\cdot I \cdot P_{\text{gen}}P_{\text{det}}\cdot D \cdot  \\ &\int _{-\infty} ^{\infty} \frac{1}{\sqrt{4 \pi Dt}}e^{-\frac{L^2}{4Dt}}\cdot e^{-\frac{t}{\tau}}cos(\Omega_L(H_\bot)t+\phi) dt. 
\end{split} \label{eq_hanle_fit_02}
\end{align}
Here, $\rho$ and $S$ are the resistivity and cross sectional area of the GaAs channel, respectively. $I$ is the applied bias current, $P_{\text{gen}}$ and $P_{\text{det}}$ are generation (injection or extraction) and detection efficiency, and $\tau$ is the spin relaxation time. $D$ is the diffusion constant which is connected to the spin diffusion length $\lambda$ by the relation ${D = \lambda^2/\tau}$. $L$ is the center-to-center distance between injection and detection strips, $\Omega_L$ is the Larmor frequency, and $\phi$ is a phase that was added to account for a random phase shift of the signals which may originate, e.g., from external stray fields. For our analysis, we used the common assumption ${P_{\text{gen}}=P_{\text{det}}}$.\cite{Ciorga2009,Bruski2013c, Manzke2017a} 
\begin{figure}[t]
\centering
\includegraphics[width=6.6cm]{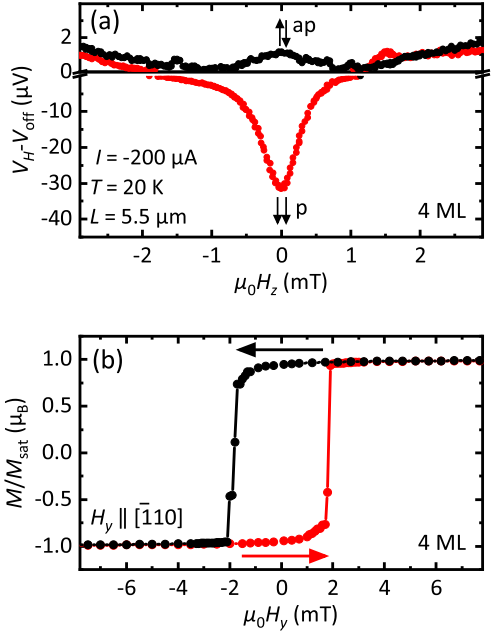}
\caption{(a)~NLH signal as a function of an external out-of-plane field measured in the parallel (p) and antiparallel (ap) strip configurations at a bias current of ${-200 \mu}$A (spin injection condition) and a temperature of 20~K. A linear background signal was subtracted. (b)~Magnetization $M$ of a CFS/MgO/GaAs structure normalized by the saturation magnetization $M_{\text{sat}}$ as a function of an external magnetic field applied along the [\={1}10] direction of the GaAs substrate (easy CFS axis) measured by SQUID magnetometry. }\label{fig5}
\end{figure}
\begin{figure}[t]
\centering
\includegraphics[width=7.5cm]{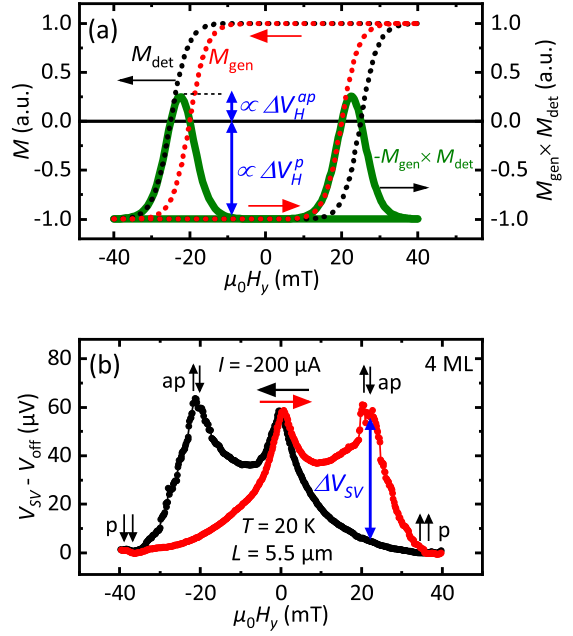}
\caption{(a) Schematic magnetization (dotted lines) of CFS generation (detection) contact $M_{\text{gen}}$ ($M_{\text{det}}$) as a function of external magnetic field. The product ${-M_{\text{gen}}\times M_{\text{det}}}$ is proportional to the resulting NLSV signal, and qualitatively predicts the non-local Hanle signal heights for parallel (${\Delta V_H^{p}}$) and antiparallel (${\Delta V_H^{ap}}$) strip configurations, indicated by  blue arrows. (b)~NLSV signal as a function of an external magnetic in-plane field applied parallel to the magnetization of injection and detection strip at a bias current of -200~$\mu$A (spin injection condition) and a temperature of 20~K. A linear background voltage was subtracted. }\label{fig6}
\end{figure}The parameters obtained by fitting the observed Hanle curves using Eq.~\ref{eq_hanle_fit_02} are given in Fig.~\ref{fig3}. For all investigated samples spin relaxation times of ${\tau \approx100}$~ns and spin diffusion lengths of ${\lambda \approx 5.6}$~$\mu$m are obtained [cf.~Fig.~\ref{fig3}(b) and (c)] which compare well with previous data reported for similar GaAs transport channels.\cite{Bruski2013c, Manzke2017a,Lou2007a} Only for small strip distances $L$ [Fig.~\ref{fig3}(a)], the finite width of the strip becomes relevant [cf.~Fig.~\ref{fig1}(a)] which commonly leads to an overestimation (underestimation) of the spin diffusion length (spin injection efficiency). As mentioned above, the large thickness of the GaAs channel hides the observation of a potential degradation caused by the atomic interdiffusion in the proximity of the MgO/GaAs interface during MBE growth.

Another method to determine $\lambda$ is given by the dependence of the amplitude of the Hanle curves at zero field ${\Delta V_H^p=V_H(B=0)-V_{\text{off}}}$ on the strip distance:\cite{Fabian,Ciorga2009,Bruski2013c} 
\begin{align}
\frac{\Delta V_{H}^p}{I}=\frac{\lambda \rho}{2S} \cdot P_{\text{gen}}P_{\text{det}}\cdot e^{-L/\lambda}.\label{eq_nlsv_distance_02}
\end{align}
Note that commonly $\Delta V_H^p$ [see Fig.~\ref{fig3}(b)] is related to the NLSV signal ${\Delta V_{SV}}$ by the expression:
\begin{align} \Delta V_{H}^p = \frac{\Delta V_{SV}}{2}, \label{eq_nlsv}
\end{align} where ${{\Delta V_{SV}}=V_{SV}(\text{ap})-V_{SV}(\text{p})}$ is defined as the difference between the SV signal in the ap and p configurations. In Fig.~\ref{fig4}, ${\vert \Delta V_H^p/I \vert}$ is shown as a function of the strip distance $L$ according to Eq.~\ref{eq_nlsv_distance_02}. Excluding structures with small contact separations because of finite strip-width effects (see discussion above), the obtained spin diffusion lengths $\lambda$ from Eq.~\ref{eq_nlsv_distance_02} given in Fig.~\ref{fig4} are in good agreement with the values obtained by fitting the whole Hanle curves to Eq.~\ref{eq_hanle_fit_02}. Because of this consistency, we rule out a significant influence of dynamic nuclear polarization on the Hanle measurements.\cite{Salis2009, Chan2009, Shiogai2012a}

In addition to the quantitative analysis of the Hanle curves in the p configuration, we also recorded Hanle curves in the ap configuration as shown in Fig.~\ref{fig5}(a) for the 4~ML MgO sample. The observed Hanle signals $\Delta V_H$ in the ap configuration are clearly smaller than those in the corresponding p configuration, as shown for the sample with 4~ML MgO in Fig.~\ref{fig5}(a). 
This finding can be explained by the following scenario: for the Hanle measurements in the p configuration, the magnetization of the CFS strips was aligned at sufficiently large external fields to ensure the magnetization of both involved CFS strips being in the saturation range.
However, a complete ap configuration may be not achievable when having non-abrupt magnetization reversal in the CFS films, as indicated by the magnetometry measurements [see Fig.~\ref{fig5}(b)] which are discussed in a separate article.\cite{Hoffmann2018a} Since the total spin polarizations in the conduction band of the FM contacts ($P_{\text{gen}}$ and $P_{\text{det}}$ in Eq.~\ref{eq_nlsv_distance_02}) is proportional to their magnetization, the resulting SV signal can be simulated by the product ${M_{\text{gen}}\times M_{\text{det}}\propto P_{\text{gen}}\times P_{\text{det}}}$ of the corresponding magnetization curves, as shown in Fig.~\ref{fig6}(a). 
When the magnetization switching of the CFS strips extends over a sufficiently large external field range, a complete ap configuration with both involved strips having saturated magnetization at the same time is never reached which leads to a reduced signal ${\Delta V_{H}^{ap} < \Delta V_H^p}$, as illustrated by the schematic magnetization curves in Fig.~\ref{fig6}(a).
Note that the product $M_{\text{gen}}\times M_{\text{det}}$ is shown in Fig.~\ref{fig6}(a) with reversed sign in order to account for the proportionality of spin signals to the bias current (Eq.~\ref{eq_nlsv_distance_02}) which has negative sign in the case of spin injection conditions.
Since the NLSV signal $V_{SV}$ is proportional to ${P_{\text{gen}}(H_y)\times P_{\text{det}}(H_y)}$ we expect two rather broad maxima for the NLSV measurements [see Fig.~\ref{fig6}(a)]. The NLSV signal for the sample with 4~ML MgO, shown in Fig.~\ref{fig6}(b), exhibits three broad maxima, from which the two maxima at ${\mu_0 H_y=\pm}$20~mT indeed resemble the expectation from our model and reach the value ${\Delta V_{SV} = \Delta V_H^p+ \Delta V_H ^{ap}}$ [see Fig.~\ref{fig6}(a)] instead of the ideal magnitude ${\Delta V_{SV} = 2\Delta V_H^p}$ according to Eq.~\ref{eq_nlsv}. In this case, the magnitude of the NLSV signals depends on the actual width of the spin-valve signals [see Fig.~\ref{fig6}(a) and discussion above]. Consequently, the stochastic nature of the switching field causes an uncertainty in the magnitude of NLSV signal for each individual measurement.\cite{Salis2010} From repeated measurements for each sample, this uncertainty can be estimated to be on the order of 10$\%$. Note that the switching fields in the NLSV measurements are larger than expected from the magnetization curves shown in Fig.~\ref{fig5}(b) as it is commonly observed,\cite{Bruski2013c,Hashimoto2006a} and most likely is a result of the shape anisotropy in the strips and other demagnetizing field effects arising from impurities at the contact edges produced during the device preparation process.\cite{Fluitman1973a,Kryder1980a,Lee1999} The large widths of the spin valve signals is accordingly attributed to the increase in the magnetization switching fields. \begin{figure}[t!]
\centering
\includegraphics[width=7.5cm]{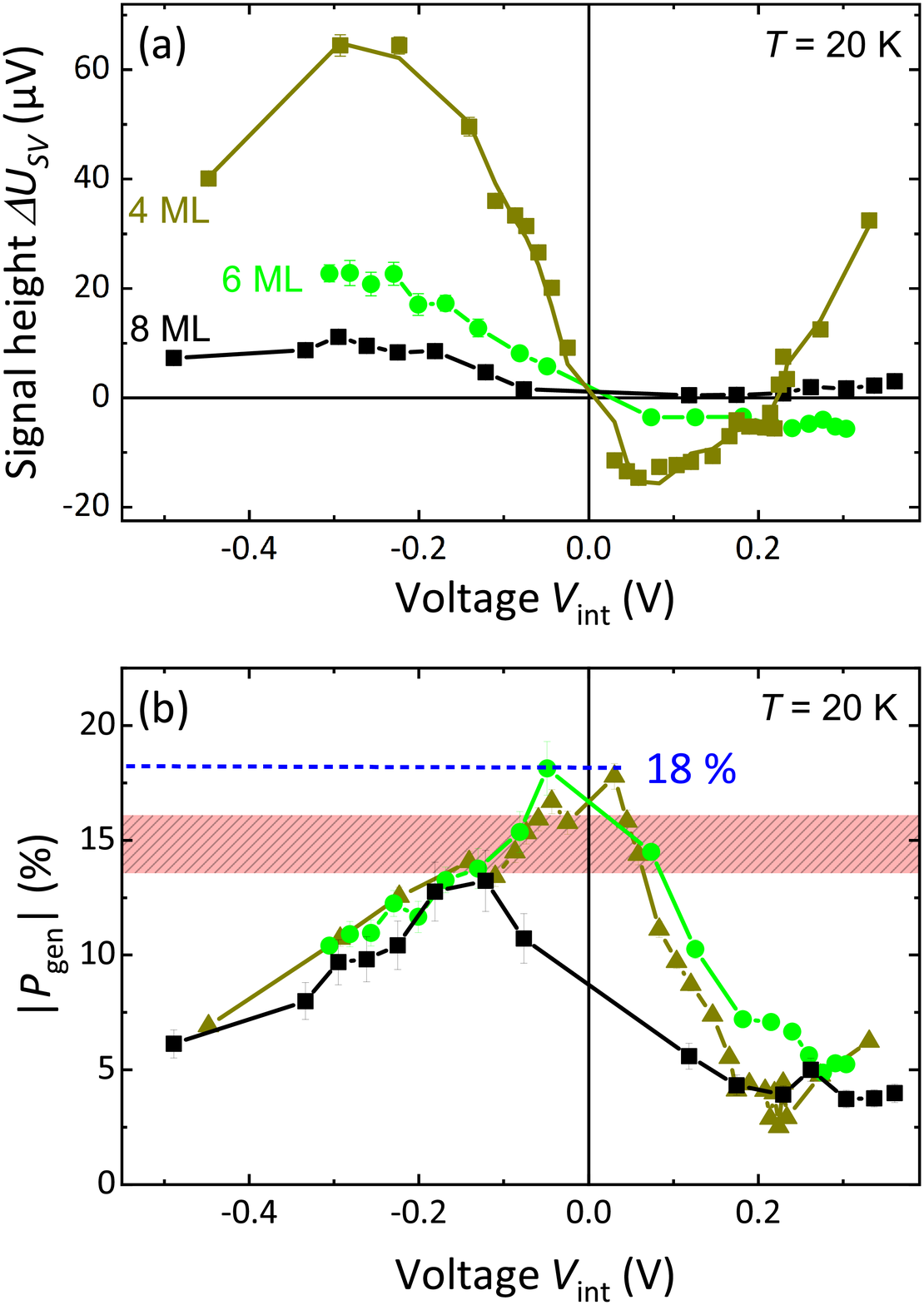}
\caption{(a) NLSV signal height determined from the measurements shown in Fig.~\ref{fig6}(b) as a function of the interface bias voltage for samples with different MgO film thicknesses. (b) spin generation efficiency $|P_{\text{gen}}| $ as a function of the interface bias voltage $V_{\text{int}}$ for the corresponding samples. The red bar indicates the range of spin injection efficiencies determined by CFS/GaAs lateral spin valves produced and measured under similar conditions.\cite{Bruski2013c}}\label{fig7}
\end{figure}
We address the third maximum around ${\mu_0 H_y = 0}$~mT to effects induced by the dynamic nuclear polarization in the GaAs channel.\cite{Salis2009, Chan2009, Shiogai2012a}

In order to gain more information on the underlying spin generation processes, we measured the non-local SV signal ${\Delta V_{SV}}$ as a function of the bias voltage $V_{int}$ for the samples with MgO shown in Fig.~\ref{fig7}(a). Since NLSV signals $\Delta V_{SV}$ are proportional to the bias current $I$,\cite{Fabian,Ciorga2009,Bruski2013c} no sign reversal is expected under injection ($I<0$) or extraction ($I>0$) conditions. For majority spin injection, positive (negative) values of $\Delta V_{SV}$ are expected for negative (positive) interface bias voltages $V_{\text{int}}$.\cite{Lou2007a} Furthermore, in the spin injection regime, and for values of ${V_{\text{int}}<-0.3}$~V, injection into the $L$ and $X$ valleys of GaAs leads to shorter spin relaxation times, and thereby reduces the SV signal.\cite{Saikin2006a,Song2010a,Salis2011a,Manzke2017a} Furthermore, spin relaxation processes related to the electric field in the FM/SC interface proximity region can lead to a bias dependent spin-injection efficiency.\cite{Valenzuela2005a,VantErve2012a} The observed decrease of the spin injection efficiency (cf. Fig.~\ref{fig7}) indeed resembles the theoretically expected and the previously experimentally observed behavior.\cite{Salis2011a,Manzke2017a} On the other hand, under spin extraction conditions, the bias dependence of the SV signal directly reflects the spin-polarized electronic band structure of CFS.\cite{Bruski2014,Manzke2017a} For the sample with 4~ML MgO in Fig.~\ref{fig7}(a), the sign of the spin signal $\Delta V_{SV}$ changes at $V_{\text{int}}=0.22$~V. This sign reversal indicates the detection of the upper edge of the halfmetallic gap in $L2_1$ CFS which has been calculated to be around 0.3~eV.\cite{Bruski2011a} However, for the other samples deviations from this ideal behavior can be observed. The bias dependent SV signals of the samples with 6~and~8~ML MgO shown in Fig.~\ref{fig7}, differ clearly from that of the sample with 4~ML MgO. In accordance with the relatively small decay constant $k$ discussed above (see Eq.~\ref{eq_buttler}), we speculate that these non-ideal behaviors can be explained by electrically active defects which influence the tunneling processes even in MBE grown CFS/MgO/GaAs contacts.\cite{Yang2011c, Guezo2008b}

In Fig.~\ref{fig7}(b), we show the spin generation efficiency $\vert P_{\text{gen}}\vert$ extracted from the data shown in Fig.~\ref{fig7}(a) using Eqs.~\ref{eq_nlsv_distance_02} and \ref{eq_nlsv} (${P_{\text{gen}} = P_{\text{det}}}$). The resulting spin generation efficiency exhibits the expected decrease with increasing $\vert V_{\text{int}}\vert$.\cite{Manzke2017a} Thereby, the bias dependences of ${P_{\text{gen}}}$ of the samples with 4~ML and 6 ML MgO are very similar. In the low bias regime, $P_{\text{gen}}$ reaches a value of $18\%$ for these samples. This value is slightly enhanced compared to the CFS/GaAs samples [see red bar in Fig.~\ref{fig7}(b)] reported earlier.\cite{Bruski2013c} Here, one has to keep in mind that the spin generation efficiencies extracted from the present NLSV measurements have to be regarded as lower limits due to the non-abrupt CFS magnetization curves [see Fig.~\ref{fig6}(a) and related discussion above]. Consequently, an improvement of the spin generation seems to be already achievable by the insertion of a MBE grown MgO interlayer, despite the non-ideal CFS/MgO/GaAs interface indicated by our results.

In conclusion, epitaxially grown MgO interlayers at the Co$_2$FeSi/GaAs interface act as tunnel barriers and allow for spin generation in the GaAs channel with a state-of-the-art efficiency. For relatively thin MgO barriers, a nearly ideal bias dependence of the spin signal can be achieved which directly reflects the characteristics of the spin-polarized band structure of Co$_2$FeSi. After further optimization regarding the abruptness of the Co$_2$FeSi magnetization curves and defect density, the hybrid system Co$_2$FeSi/MgO/GaAs will be a very promising spintronic building block, free of electrical compensation due to atomic interdiffusion during film growth.

\section{Acknowledgments}\label{sec_ackn}
The authors thank C.~Herrmann and H.-P.~Sch\"{o}nherr for their valuable support during the growth of the samples as well as Walid Anders  for the chemical preparation, Angela Riedel for support during the electrical measurements, Samuel Gaucher and Stefan Ludwig for helping with the manuscript.\\


\begin{thebibliography}{0}%
\makeatletter
\providecommand \@ifxundefined [1]{%
 \@ifx{#1\undefined}
}%
\providecommand \@ifnum [1]{%
 \ifnum #1\expandafter \@firstoftwo
 \else \expandafter \@secondoftwo
 \fi
}%
\providecommand \@ifx [1]{%
 \ifx #1\expandafter \@firstoftwo
 \else \expandafter \@secondoftwo
 \fi
}%
\providecommand \natexlab [1]{#1}%
\providecommand \enquote  [1]{``#1''}%
\providecommand \bibnamefont  [1]{#1}%
\providecommand \bibfnamefont [1]{#1}%
\providecommand \citenamefont [1]{#1}%
\providecommand \href@noop [0]{\@secondoftwo}%
\providecommand \href [0]{\begingroup \@sanitize@url \@href}%
\providecommand \@href[1]{\@@startlink{#1}\@@href}%
\providecommand \@@href[1]{\endgroup#1\@@endlink}%
\providecommand \@sanitize@url [0]{\catcode `\\12\catcode `\$12\catcode
  `\&12\catcode `\#12\catcode `\^12\catcode `\_12\catcode `\%12\relax}%
\providecommand \@@startlink[1]{}%
\providecommand \@@endlink[0]{}%
\providecommand \url  [0]{\begingroup\@sanitize@url \@url }%
\providecommand \@url [1]{\endgroup\@href {#1}{\urlprefix }}%
\providecommand \urlprefix  [0]{URL }%
\providecommand \Eprint [0]{\href }%
\providecommand \doibase [0]{http://dx.doi.org/}%
\providecommand \selectlanguage [0]{\@gobble}%
\providecommand \bibinfo  [0]{\@secondoftwo}%
\providecommand \bibfield  [0]{\@secondoftwo}%
\providecommand \translation [1]{[#1]}%
\providecommand \BibitemOpen [0]{}%
\providecommand \bibitemStop [0]{}%
\providecommand \bibitemNoStop [0]{.\EOS\space}%
\providecommand \EOS [0]{\spacefactor3000\relax}%
\providecommand \BibitemShut  [1]{\csname bibitem#1\endcsname}%
\let\auto@bib@innerbib\@empty
\end{thebibliography}%


\begin{thebibliography}{60}%
\makeatletter
\providecommand \@ifxundefined [1]{%
 \@ifx{#1\undefined}
}%
\providecommand \@ifnum [1]{%
 \ifnum #1\expandafter \@firstoftwo
 \else \expandafter \@secondoftwo
 \fi
}%
\providecommand \@ifx [1]{%
 \ifx #1\expandafter \@firstoftwo
 \else \expandafter \@secondoftwo
 \fi
}%
\providecommand \natexlab [1]{#1}%
\providecommand \enquote  [1]{``#1''}%
\providecommand \bibnamefont  [1]{#1}%
\providecommand \bibfnamefont [1]{#1}%
\providecommand \citenamefont [1]{#1}%
\providecommand \href@noop [0]{\@secondoftwo}%
\providecommand \href [0]{\begingroup \@sanitize@url \@href}%
\providecommand \@href[1]{\@@startlink{#1}\@@href}%
\providecommand \@@href[1]{\endgroup#1\@@endlink}%
\providecommand \@sanitize@url [0]{\catcode `\\12\catcode `\$12\catcode
  `\&12\catcode `\#12\catcode `\^12\catcode `\_12\catcode `\%12\relax}%
\providecommand \@@startlink[1]{}%
\providecommand \@@endlink[0]{}%
\providecommand \url  [0]{\begingroup\@sanitize@url \@url }%
\providecommand \@url [1]{\endgroup\@href {#1}{\urlprefix }}%
\providecommand \urlprefix  [0]{URL }%
\providecommand \Eprint [0]{\href }%
\providecommand \doibase [0]{http://dx.doi.org/}%
\providecommand \selectlanguage [0]{\@gobble}%
\providecommand \bibinfo  [0]{\@secondoftwo}%
\providecommand \bibfield  [0]{\@secondoftwo}%
\providecommand \translation [1]{[#1]}%
\providecommand \BibitemOpen [0]{}%
\providecommand \bibitemStop [0]{}%
\providecommand \bibitemNoStop [0]{.\EOS\space}%
\providecommand \EOS [0]{\spacefactor3000\relax}%
\providecommand \BibitemShut  [1]{\csname bibitem#1\endcsname}%
\let\auto@bib@innerbib\@empty
\bibitem [{\citenamefont {Zutic}\ \emph {et~al.}(2004)\citenamefont {Zutic},
  \citenamefont {Fabian},\ and\ \citenamefont {Sarma}}]{Fabian2004}%
  \BibitemOpen
  \bibfield  {author} {\bibinfo {author} {\bibfnamefont {I.}~\bibnamefont
  {Zutic}}, \bibinfo {author} {\bibfnamefont {J.}~\bibnamefont {Fabian}}, \
  and\ \bibinfo {author} {\bibfnamefont {S.~D.}\ \bibnamefont {Sarma}},\
  }\href@noop {} {\bibfield  {journal} {\bibinfo  {journal} {Rev. Mod. Phys.}\
  }\textbf {\bibinfo {volume} {76}},\ \bibinfo {pages} {323} (\bibinfo {year}
  {2004})}\BibitemShut {NoStop}%
\bibitem [{\citenamefont {Datta}\ and\ \citenamefont {Das}(1990)}]{Datta1990a}%
  \BibitemOpen
  \bibfield  {author} {\bibinfo {author} {\bibfnamefont {S.}~\bibnamefont
  {Datta}}\ and\ \bibinfo {author} {\bibfnamefont {B.}~\bibnamefont {Das}},\
  }\href {\doibase 10.1063/1.102730} {\bibfield  {journal} {\bibinfo  {journal}
  {Appl. Phys. Lett.}\ }\textbf {\bibinfo {volume} {56}},\ \bibinfo {pages}
  {665} (\bibinfo {year} {1990})}\BibitemShut {NoStop}%
\bibitem [{\citenamefont {Wurmehl}\ \emph {et~al.}(2005)\citenamefont
  {Wurmehl}, \citenamefont {Fecher}, \citenamefont {Kandpal}, \citenamefont
  {Ksenofontov}, \citenamefont {Felser}, \citenamefont {Lin},\ and\
  \citenamefont {Morais}}]{Wurmehl2005}%
  \BibitemOpen
  \bibfield  {author} {\bibinfo {author} {\bibfnamefont {S.}~\bibnamefont
  {Wurmehl}}, \bibinfo {author} {\bibfnamefont {G.~H.}\ \bibnamefont {Fecher}},
  \bibinfo {author} {\bibfnamefont {H.~C.}\ \bibnamefont {Kandpal}}, \bibinfo
  {author} {\bibfnamefont {V.}~\bibnamefont {Ksenofontov}}, \bibinfo {author}
  {\bibfnamefont {C.}~\bibnamefont {Felser}}, \bibinfo {author} {\bibfnamefont
  {H.-J.}\ \bibnamefont {Lin}}, \ and\ \bibinfo {author} {\bibfnamefont
  {J.}~\bibnamefont {Morais}},\ }\href {\doibase 10.1103/PhysRevB.72.184434}
  {\bibfield  {journal} {\bibinfo  {journal} {Phys. Rev. B}\ }\textbf {\bibinfo
  {volume} {72}},\ \bibinfo {pages} {184434} (\bibinfo {year}
  {2005})}\BibitemShut {NoStop}%
\bibitem [{\citenamefont {Manzke}\ \emph {et~al.}(2013)\citenamefont {Manzke},
  \citenamefont {Farshchi}, \citenamefont {Bruski}, \citenamefont {Herfort},\
  and\ \citenamefont {Ramsteiner}}]{Manzke2013}%
  \BibitemOpen
  \bibfield  {author} {\bibinfo {author} {\bibfnamefont {Y.}~\bibnamefont
  {Manzke}}, \bibinfo {author} {\bibfnamefont {R.}~\bibnamefont {Farshchi}},
  \bibinfo {author} {\bibfnamefont {P.}~\bibnamefont {Bruski}}, \bibinfo
  {author} {\bibfnamefont {J.}~\bibnamefont {Herfort}}, \ and\ \bibinfo
  {author} {\bibfnamefont {M.}~\bibnamefont {Ramsteiner}},\ }\href {\doibase
  10.1103/PhysRevB.87.134415} {\bibfield  {journal} {\bibinfo  {journal} {Phys.
  Rev. B}\ }\textbf {\bibinfo {volume} {87}},\ \bibinfo {pages} {134415}
  (\bibinfo {year} {2013})}\BibitemShut {NoStop}%
\bibitem [{\citenamefont {Crooker}\ \emph {et~al.}(2005)\citenamefont
  {Crooker}, \citenamefont {Furis}, \citenamefont {Lou}, \citenamefont
  {Adelmann}, \citenamefont {Smith}, \citenamefont {Palmstr{\o}m},\ and\
  \citenamefont {Crowell}}]{Crooker2005a}%
  \BibitemOpen
  \bibfield  {author} {\bibinfo {author} {\bibfnamefont {S.~A.}\ \bibnamefont
  {Crooker}}, \bibinfo {author} {\bibfnamefont {M.}~\bibnamefont {Furis}},
  \bibinfo {author} {\bibfnamefont {X.}~\bibnamefont {Lou}}, \bibinfo {author}
  {\bibfnamefont {C.}~\bibnamefont {Adelmann}}, \bibinfo {author}
  {\bibfnamefont {D.~L.}\ \bibnamefont {Smith}}, \bibinfo {author}
  {\bibfnamefont {C.~J.}\ \bibnamefont {Palmstr{\o}m}}, \ and\ \bibinfo
  {author} {\bibfnamefont {P.~A.}\ \bibnamefont {Crowell}},\ }\href {\doibase
  10.1126/science.1116865} {\bibfield  {journal} {\bibinfo  {journal}
  {Science}\ }\textbf {\bibinfo {volume} {309}},\ \bibinfo {pages} {2191}
  (\bibinfo {year} {2005})}\BibitemShut {NoStop}%
\bibitem [{\citenamefont {Peterson}\ \emph {et~al.}(2016)\citenamefont
  {Peterson}, \citenamefont {Patel}, \citenamefont {Geppert}, \citenamefont
  {Christie}, \citenamefont {Rath}, \citenamefont {Pennachio}, \citenamefont
  {Flatt{\'{e}}}, \citenamefont {Voyles}, \citenamefont {Palmstr{\o}m},\ and\
  \citenamefont {Crowell}}]{Peterson2016}%
  \BibitemOpen
  \bibfield  {author} {\bibinfo {author} {\bibfnamefont {T.~A.}\ \bibnamefont
  {Peterson}}, \bibinfo {author} {\bibfnamefont {S.~J.}\ \bibnamefont {Patel}},
  \bibinfo {author} {\bibfnamefont {C.~C.}\ \bibnamefont {Geppert}}, \bibinfo
  {author} {\bibfnamefont {K.~D.}\ \bibnamefont {Christie}}, \bibinfo {author}
  {\bibfnamefont {A.}~\bibnamefont {Rath}}, \bibinfo {author} {\bibfnamefont
  {D.}~\bibnamefont {Pennachio}}, \bibinfo {author} {\bibfnamefont {M.~E.}\
  \bibnamefont {Flatt{\'{e}}}}, \bibinfo {author} {\bibfnamefont {P.~M.}\
  \bibnamefont {Voyles}}, \bibinfo {author} {\bibfnamefont {C.~J.}\
  \bibnamefont {Palmstr{\o}m}}, \ and\ \bibinfo {author} {\bibfnamefont
  {P.~A.}\ \bibnamefont {Crowell}},\ }\href {\doibase
  10.1103/PhysRevB.94.235309} {\bibfield  {journal} {\bibinfo  {journal} {Phys.
  Rev. B}\ }\textbf {\bibinfo {volume} {94}},\ \bibinfo {pages} {235309}
  (\bibinfo {year} {2016})}\BibitemShut {NoStop}%
\bibitem [{\citenamefont {Webster}(1969)}]{Webster1969}%
  \BibitemOpen
  \bibfield  {author} {\bibinfo {author} {\bibfnamefont {P.~J.}\ \bibnamefont
  {Webster}},\ }\href {\doibase 10.1080/00107516908204800} {\bibfield
  {journal} {\bibinfo  {journal} {Contemp. Phys.}\ }\textbf {\bibinfo {volume}
  {10}},\ \bibinfo {pages} {559} (\bibinfo {year} {1969})}\BibitemShut
  {NoStop}%
\bibitem [{\citenamefont {Hu}\ \emph {et~al.}(2011)\citenamefont {Hu},
  \citenamefont {Garlid}, \citenamefont {Crowell},\ and\ \citenamefont
  {Palmstr{\o}m}}]{Hu2011b}%
  \BibitemOpen
  \bibfield  {author} {\bibinfo {author} {\bibfnamefont {Q.~O.}\ \bibnamefont
  {Hu}}, \bibinfo {author} {\bibfnamefont {E.~S.}\ \bibnamefont {Garlid}},
  \bibinfo {author} {\bibfnamefont {P.~A.}\ \bibnamefont {Crowell}}, \ and\
  \bibinfo {author} {\bibfnamefont {C.~J.}\ \bibnamefont {Palmstr{\o}m}},\
  }\href {\doibase 10.1103/PhysRevB.84.085306} {\bibfield  {journal} {\bibinfo
  {journal} {Phys. Rev. B}\ }\textbf {\bibinfo {volume} {84}},\ \bibinfo
  {pages} {085306} (\bibinfo {year} {2011})}\BibitemShut {NoStop}%
\bibitem [{\citenamefont {Manzke}\ \emph {et~al.}(2017)\citenamefont {Manzke},
  \citenamefont {Herfort},\ and\ \citenamefont {Ramsteiner}}]{Manzke2017a}%
  \BibitemOpen
  \bibfield  {author} {\bibinfo {author} {\bibfnamefont {Y.}~\bibnamefont
  {Manzke}}, \bibinfo {author} {\bibfnamefont {J.}~\bibnamefont {Herfort}}, \
  and\ \bibinfo {author} {\bibfnamefont {M.}~\bibnamefont {Ramsteiner}},\
  }\href {\doibase 10.1103/PhysRevB.96.245308} {\bibfield  {journal} {\bibinfo
  {journal} {Phys. Rev. B}\ }\textbf {\bibinfo {volume} {96}},\ \bibinfo
  {pages} {245308} (\bibinfo {year} {2017})}\BibitemShut {NoStop}%
\bibitem [{\citenamefont {Brandt}\ \emph {et~al.}(2010)\citenamefont {Brandt},
  \citenamefont {Ramsteiner}, \citenamefont {Flissikowski}, \citenamefont
  {Herfort},\ and\ \citenamefont {Grahn}}]{Brandt2010}%
  \BibitemOpen
  \bibfield  {author} {\bibinfo {author} {\bibfnamefont {O.}~\bibnamefont
  {Brandt}}, \bibinfo {author} {\bibfnamefont {M.}~\bibnamefont {Ramsteiner}},
  \bibinfo {author} {\bibfnamefont {T.}~\bibnamefont {Flissikowski}}, \bibinfo
  {author} {\bibfnamefont {J.}~\bibnamefont {Herfort}}, \ and\ \bibinfo
  {author} {\bibfnamefont {H.~T.}\ \bibnamefont {Grahn}},\ }\href {\doibase
  10.1103/PhysRevB.81.115302} {\bibfield  {journal} {\bibinfo  {journal} {Phys.
  Rev. B}\ }\textbf {\bibinfo {volume} {81}},\ \bibinfo {pages} {115302}
  (\bibinfo {year} {2010})}\BibitemShut {NoStop}%
\bibitem [{\citenamefont {Hashimoto}\ \emph {et~al.}(2007)\citenamefont
  {Hashimoto}, \citenamefont {Trampert}, \citenamefont {Herfort},\ and\
  \citenamefont {Ploog}}]{Hashimoto2007b}%
  \BibitemOpen
  \bibfield  {author} {\bibinfo {author} {\bibfnamefont {M.}~\bibnamefont
  {Hashimoto}}, \bibinfo {author} {\bibfnamefont {A.}~\bibnamefont {Trampert}},
  \bibinfo {author} {\bibfnamefont {J.}~\bibnamefont {Herfort}}, \ and\
  \bibinfo {author} {\bibfnamefont {K.~H.}\ \bibnamefont {Ploog}},\ }\href
  {\doibase 10.1116/1.2748413} {\bibfield  {journal} {\bibinfo  {journal} {J.
  Vac. Sci. Technol. B}\ }\textbf {\bibinfo {volume} {25}},\ \bibinfo {pages}
  {1453} (\bibinfo {year} {2007})}\BibitemShut {NoStop}%
\bibitem [{\citenamefont {Jenichen}\ \emph {et~al.}(2012)\citenamefont
  {Jenichen}, \citenamefont {Herfort}, \citenamefont {Hentschel}, \citenamefont
  {Nikulin}, \citenamefont {Kong}, \citenamefont {Trampert},\ and\
  \citenamefont {Aiak}}]{Jenichen2012a}%
  \BibitemOpen
  \bibfield  {author} {\bibinfo {author} {\bibfnamefont {B.}~\bibnamefont
  {Jenichen}}, \bibinfo {author} {\bibfnamefont {J.}~\bibnamefont {Herfort}},
  \bibinfo {author} {\bibfnamefont {T.}~\bibnamefont {Hentschel}}, \bibinfo
  {author} {\bibfnamefont {A.}~\bibnamefont {Nikulin}}, \bibinfo {author}
  {\bibfnamefont {X.}~\bibnamefont {Kong}}, \bibinfo {author} {\bibfnamefont
  {A.}~\bibnamefont {Trampert}}, \ and\ \bibinfo {author} {\bibfnamefont
  {I.}~\bibnamefont {Aiak}},\ }\href {\doibase 10.1103/PhysRevB.86.075319}
  {\bibfield  {journal} {\bibinfo  {journal} {Phys. Rev. B}\ }\textbf {\bibinfo
  {volume} {86}},\ \bibinfo {pages} {075319} (\bibinfo {year}
  {2012})}\BibitemShut {NoStop}%
\bibitem [{\citenamefont {Hentschel}(2015)}]{Hentschel2015a}%
  \BibitemOpen
  \bibfield  {author} {\bibinfo {author} {\bibfnamefont {T.}~\bibnamefont
  {Hentschel}},\ }\emph {\bibinfo {title} {{Wachstum und Charakterisierung von
  Seltenerdoxiden und Magnesiumoxid auf Galliumarsenid-Substraten: Diffusions-
  und Tunnelbarrieren in Ferromagnet/Halbleiter-Hybridstrukturen}}},\
  \href@noop {} {Ph.D. thesis},\ \bibinfo  {school} {Humboldt-Universit{\"{a}}t
  zu Berlin} (\bibinfo {year} {2015})\BibitemShut {NoStop}%
\bibitem [{\citenamefont {Makarov}\ \emph {et~al.}(2011)\citenamefont
  {Makarov}, \citenamefont {Krumme}, \citenamefont {Stromberg}, \citenamefont
  {Weis}, \citenamefont {Keune},\ and\ \citenamefont {Wende}}]{Makarov2011b}%
  \BibitemOpen
  \bibfield  {author} {\bibinfo {author} {\bibfnamefont {S.~I.}\ \bibnamefont
  {Makarov}}, \bibinfo {author} {\bibfnamefont {B.}~\bibnamefont {Krumme}},
  \bibinfo {author} {\bibfnamefont {F.}~\bibnamefont {Stromberg}}, \bibinfo
  {author} {\bibfnamefont {C.}~\bibnamefont {Weis}}, \bibinfo {author}
  {\bibfnamefont {W.}~\bibnamefont {Keune}}, \ and\ \bibinfo {author}
  {\bibfnamefont {H.}~\bibnamefont {Wende}},\ }\href {\doibase
  10.1063/1.3646390} {\bibfield  {journal} {\bibinfo  {journal} {Appl. Phys.
  Lett.}\ }\textbf {\bibinfo {volume} {99}},\ \bibinfo {pages} {141910}
  (\bibinfo {year} {2011})}\BibitemShut {NoStop}%
\bibitem [{\citenamefont {Butler}\ \emph {et~al.}(2005)\citenamefont {Butler},
  \citenamefont {Zhang}, \citenamefont {Vutukuri}, \citenamefont {Chshievand},\
  and\ \citenamefont {Schulthess}}]{Butler2005}%
  \BibitemOpen
  \bibfield  {author} {\bibinfo {author} {\bibfnamefont {W.~H.}\ \bibnamefont
  {Butler}}, \bibinfo {author} {\bibfnamefont {X.~G.}\ \bibnamefont {Zhang}},
  \bibinfo {author} {\bibfnamefont {S.}~\bibnamefont {Vutukuri}}, \bibinfo
  {author} {\bibfnamefont {M.}~\bibnamefont {Chshievand}}, \ and\ \bibinfo
  {author} {\bibfnamefont {T.~G.}\ \bibnamefont {Schulthess}},\ }\href
  {\doibase 10.1109/TMAG.2005.854763} {\bibfield  {journal} {\bibinfo
  {journal} {IEEE Trans. Magn.}\ }\textbf {\bibinfo {volume} {41}},\ \bibinfo
  {pages} {2645} (\bibinfo {year} {2005})}\BibitemShut {NoStop}%
\bibitem [{\citenamefont {Butler}(2008)}]{Butler2008}%
  \BibitemOpen
  \bibfield  {author} {\bibinfo {author} {\bibfnamefont {W.~H.}\ \bibnamefont
  {Butler}},\ }\href {\doibase 10.1088/1468-6996/9/1/014106} {\bibfield
  {journal} {\bibinfo  {journal} {Sci. Technol. Adv. Mater.}\ }\textbf
  {\bibinfo {volume} {9}},\ \bibinfo {pages} {014106} (\bibinfo {year}
  {2008})}\BibitemShut {NoStop}%
\bibitem [{\citenamefont {Yang}\ \emph {et~al.}(2011)\citenamefont {Yang},
  \citenamefont {Balke}, \citenamefont {Papp}, \citenamefont {D{\"{o}}ring},
  \citenamefont {Berges}, \citenamefont {Plucinski}, \citenamefont {Westphal},
  \citenamefont {Schneider}, \citenamefont {Parkin},\ and\ \citenamefont
  {Fadley}}]{Yang2011c}%
  \BibitemOpen
  \bibfield  {author} {\bibinfo {author} {\bibfnamefont {S.~H.}\ \bibnamefont
  {Yang}}, \bibinfo {author} {\bibfnamefont {B.}~\bibnamefont {Balke}},
  \bibinfo {author} {\bibfnamefont {C.}~\bibnamefont {Papp}}, \bibinfo {author}
  {\bibfnamefont {S.}~\bibnamefont {D{\"{o}}ring}}, \bibinfo {author}
  {\bibfnamefont {U.}~\bibnamefont {Berges}}, \bibinfo {author} {\bibfnamefont
  {L.}~\bibnamefont {Plucinski}}, \bibinfo {author} {\bibfnamefont
  {C.}~\bibnamefont {Westphal}}, \bibinfo {author} {\bibfnamefont {C.~M.}\
  \bibnamefont {Schneider}}, \bibinfo {author} {\bibfnamefont {S.~S.}\
  \bibnamefont {Parkin}}, \ and\ \bibinfo {author} {\bibfnamefont {C.~S.}\
  \bibnamefont {Fadley}},\ }\href {\doibase 10.1103/PhysRevB.84.184410}
  {\bibfield  {journal} {\bibinfo  {journal} {Phys. Rev. B}\ }\textbf {\bibinfo
  {volume} {84}},\ \bibinfo {pages} {184410} (\bibinfo {year}
  {2011})}\BibitemShut {NoStop}%
\bibitem [{\citenamefont {Gu{\'{e}}zo}\ \emph {et~al.}(2008)\citenamefont
  {Gu{\'{e}}zo}, \citenamefont {Turban}, \citenamefont {Lallaizon},
  \citenamefont {{Le Breton}}, \citenamefont {Schieffer}, \citenamefont
  {L{\'{e}}pine},\ and\ \citenamefont {J{\'{e}}z{\'{e}}quel}}]{Guezo2008b}%
  \BibitemOpen
  \bibfield  {author} {\bibinfo {author} {\bibfnamefont {S.}~\bibnamefont
  {Gu{\'{e}}zo}}, \bibinfo {author} {\bibfnamefont {P.}~\bibnamefont {Turban}},
  \bibinfo {author} {\bibfnamefont {C.}~\bibnamefont {Lallaizon}}, \bibinfo
  {author} {\bibfnamefont {J.~C.}\ \bibnamefont {{Le Breton}}}, \bibinfo
  {author} {\bibfnamefont {P.}~\bibnamefont {Schieffer}}, \bibinfo {author}
  {\bibfnamefont {B.}~\bibnamefont {L{\'{e}}pine}}, \ and\ \bibinfo {author}
  {\bibfnamefont {G.}~\bibnamefont {J{\'{e}}z{\'{e}}quel}},\ }\href {\doibase
  10.1063/1.3012571} {\bibfield  {journal} {\bibinfo  {journal} {Appl. Phys.
  Lett.}\ }\textbf {\bibinfo {volume} {93}},\ \bibinfo {pages} {172116}
  (\bibinfo {year} {2008})}\BibitemShut {NoStop}%
\bibitem [{\citenamefont {Yuasa}\ \emph {et~al.}(2004)\citenamefont {Yuasa},
  \citenamefont {Nagahama}, \citenamefont {Fukushima}, \citenamefont {Suzuki},\
  and\ \citenamefont {Ando}}]{Yuasa2004}%
  \BibitemOpen
  \bibfield  {author} {\bibinfo {author} {\bibfnamefont {S.}~\bibnamefont
  {Yuasa}}, \bibinfo {author} {\bibfnamefont {T.}~\bibnamefont {Nagahama}},
  \bibinfo {author} {\bibfnamefont {A.}~\bibnamefont {Fukushima}}, \bibinfo
  {author} {\bibfnamefont {Y.}~\bibnamefont {Suzuki}}, \ and\ \bibinfo {author}
  {\bibfnamefont {K.}~\bibnamefont {Ando}},\ }\href {\doibase 10.1038/nmat1257}
  {\bibfield  {journal} {\bibinfo  {journal} {Nat. Mater.}\ }\textbf {\bibinfo
  {volume} {3}},\ \bibinfo {pages} {868} (\bibinfo {year} {2004})}\BibitemShut
  {NoStop}%
\bibitem [{\citenamefont {Hoffmann}\ \emph {et~al.}(2019)\citenamefont
  {Hoffmann}, \citenamefont {Jenichen},\ and\ \citenamefont
  {Herfort}}]{Hoffmann2018a}%
  \BibitemOpen
  \bibfield  {author} {\bibinfo {author} {\bibfnamefont {G.}~\bibnamefont
  {Hoffmann}}, \bibinfo {author} {\bibfnamefont {B.}~\bibnamefont {Jenichen}},
  \ and\ \bibinfo {author} {\bibfnamefont {J.}~\bibnamefont {Herfort}},\ }\href
  {\doibase 10.1016/j.jcrysgro.2019.02.029} {\bibfield  {journal} {\bibinfo
  {journal} {J. Cryst. Growth}\ }\textbf {\bibinfo {volume} {512}},\ \bibinfo
  {pages} {194} (\bibinfo {year} {2019})}\BibitemShut {NoStop}%
\bibitem [{\citenamefont {Bruski}\ \emph {et~al.}(2013)\citenamefont {Bruski},
  \citenamefont {Manzke}, \citenamefont {Farshchi}, \citenamefont {Brandt},
  \citenamefont {Herfort},\ and\ \citenamefont {Ramsteiner}}]{Bruski2013c}%
  \BibitemOpen
  \bibfield  {author} {\bibinfo {author} {\bibfnamefont {P.}~\bibnamefont
  {Bruski}}, \bibinfo {author} {\bibfnamefont {Y.}~\bibnamefont {Manzke}},
  \bibinfo {author} {\bibfnamefont {R.}~\bibnamefont {Farshchi}}, \bibinfo
  {author} {\bibfnamefont {O.}~\bibnamefont {Brandt}}, \bibinfo {author}
  {\bibfnamefont {J.}~\bibnamefont {Herfort}}, \ and\ \bibinfo {author}
  {\bibfnamefont {M.}~\bibnamefont {Ramsteiner}},\ }\href {\doibase
  10.1063/1.4817270} {\bibfield  {journal} {\bibinfo  {journal} {Appl. Phys.
  Lett.}\ }\textbf {\bibinfo {volume} {103}},\ \bibinfo {pages} {052406}
  (\bibinfo {year} {2013})}\BibitemShut {NoStop}%
\bibitem [{\citenamefont {Uemura}\ \emph {et~al.}(2012)\citenamefont {Uemura},
  \citenamefont {Kondo}, \citenamefont {Fujisawa}, \citenamefont {Matsuda},\
  and\ \citenamefont {Yamamoto}}]{Uemura2012c}%
  \BibitemOpen
  \bibfield  {author} {\bibinfo {author} {\bibfnamefont {T.}~\bibnamefont
  {Uemura}}, \bibinfo {author} {\bibfnamefont {K.}~\bibnamefont {Kondo}},
  \bibinfo {author} {\bibfnamefont {J.}~\bibnamefont {Fujisawa}}, \bibinfo
  {author} {\bibfnamefont {K.~I.}\ \bibnamefont {Matsuda}}, \ and\ \bibinfo
  {author} {\bibfnamefont {M.}~\bibnamefont {Yamamoto}},\ }\href {\doibase
  10.1063/1.4754545} {\bibfield  {journal} {\bibinfo  {journal} {Appl. Phys.
  Lett.}\ }\textbf {\bibinfo {volume} {101}},\ \bibinfo {pages} {132411}
  (\bibinfo {year} {2012})}\BibitemShut {NoStop}%
\bibitem [{\citenamefont {Rashba}(2000)}]{Rashba2000}%
  \BibitemOpen
  \bibfield  {author} {\bibinfo {author} {\bibfnamefont {E.~I.}\ \bibnamefont
  {Rashba}},\ }\href {\doibase 10.1103/PhysRevB.62.R16267} {\bibfield
  {journal} {\bibinfo  {journal} {Phys. Rev. B}\ }\textbf {\bibinfo {volume}
  {62}},\ \bibinfo {pages} {267} (\bibinfo {year} {2000})}\BibitemShut
  {NoStop}%
\bibitem [{\citenamefont {Fert}\ \emph {et~al.}(2007)\citenamefont {Fert},
  \citenamefont {George}, \citenamefont {Jaffr{\`{e}}s},\ and\ \citenamefont
  {Mattana}}]{Fert2007}%
  \BibitemOpen
  \bibfield  {author} {\bibinfo {author} {\bibfnamefont {A.}~\bibnamefont
  {Fert}}, \bibinfo {author} {\bibfnamefont {J.}~\bibnamefont {George}},
  \bibinfo {author} {\bibfnamefont {H.}~\bibnamefont {Jaffr{\`{e}}s}}, \ and\
  \bibinfo {author} {\bibfnamefont {R.}~\bibnamefont {Mattana}},\ }\href
  {\doibase 10.1109/TED.2007.894372} {\bibfield  {journal} {\bibinfo  {journal}
  {IEEE Trans. Electron Devices}\ }\textbf {\bibinfo {volume} {54}},\ \bibinfo
  {pages} {921} (\bibinfo {year} {2007})}\BibitemShut {NoStop}%
\bibitem [{\citenamefont {Hanbicki}\ \emph {et~al.}(2003)\citenamefont
  {Hanbicki}, \citenamefont {{Van't Erve}}, \citenamefont {Magno},
  \citenamefont {Kioseoglou}, \citenamefont {Li}, \citenamefont {Jonker},
  \citenamefont {Itskos}, \citenamefont {Mallory}, \citenamefont {Yasar},\ and\
  \citenamefont {Petrou}}]{Hanbicki2003a}%
  \BibitemOpen
  \bibfield  {author} {\bibinfo {author} {\bibfnamefont {A.~T.}\ \bibnamefont
  {Hanbicki}}, \bibinfo {author} {\bibfnamefont {O.~M.~J.}\ \bibnamefont
  {{Van't Erve}}}, \bibinfo {author} {\bibfnamefont {R.}~\bibnamefont {Magno}},
  \bibinfo {author} {\bibfnamefont {G.}~\bibnamefont {Kioseoglou}}, \bibinfo
  {author} {\bibfnamefont {C.~H.}\ \bibnamefont {Li}}, \bibinfo {author}
  {\bibfnamefont {B.~T.}\ \bibnamefont {Jonker}}, \bibinfo {author}
  {\bibfnamefont {G.}~\bibnamefont {Itskos}}, \bibinfo {author} {\bibfnamefont
  {R.}~\bibnamefont {Mallory}}, \bibinfo {author} {\bibfnamefont
  {M.}~\bibnamefont {Yasar}}, \ and\ \bibinfo {author} {\bibfnamefont
  {A.}~\bibnamefont {Petrou}},\ }\href {\doibase 10.1063/1.1580631} {\bibfield
  {journal} {\bibinfo  {journal} {Appl. Phys. Lett.}\ }\textbf {\bibinfo
  {volume} {82}},\ \bibinfo {pages} {4092} (\bibinfo {year}
  {2003})}\BibitemShut {NoStop}%
\bibitem [{\citenamefont {J{\"{o}}nsson-{\AA}kerman}\ \emph
  {et~al.}(2002)\citenamefont {J{\"{o}}nsson-{\AA}kerman}, \citenamefont
  {Rabson}, \citenamefont {Leighton}, \citenamefont {Schuller}, \citenamefont
  {Kim},\ and\ \citenamefont {Escudero}}]{Jonsson-Akerman2002}%
  \BibitemOpen
  \bibfield  {author} {\bibinfo {author} {\bibfnamefont {B.~J.}\ \bibnamefont
  {J{\"{o}}nsson-{\AA}kerman}}, \bibinfo {author} {\bibfnamefont {D.~A.}\
  \bibnamefont {Rabson}}, \bibinfo {author} {\bibfnamefont {C.}~\bibnamefont
  {Leighton}}, \bibinfo {author} {\bibfnamefont {I.~K.}\ \bibnamefont
  {Schuller}}, \bibinfo {author} {\bibfnamefont {S.}~\bibnamefont {Kim}}, \
  and\ \bibinfo {author} {\bibfnamefont {R.}~\bibnamefont {Escudero}},\ }\href
  {\doibase 10.1063/1.1310633} {\bibfield  {journal} {\bibinfo  {journal}
  {Appl. Phys. Lett.}\ }\textbf {\bibinfo {volume} {77}},\ \bibinfo {pages}
  {1870} (\bibinfo {year} {2002})}\BibitemShut {NoStop}%
\bibitem [{\citenamefont {Simmons}(1963)}]{Simmons1963}%
  \BibitemOpen
  \bibfield  {author} {\bibinfo {author} {\bibfnamefont {J.~G.}\ \bibnamefont
  {Simmons}},\ }\href {\doibase 10.1063/1.1702682} {\bibfield  {journal}
  {\bibinfo  {journal} {J. Appl. Phys.}\ }\textbf {\bibinfo {volume} {34}},\
  \bibinfo {pages} {1793} (\bibinfo {year} {1963})}\BibitemShut {NoStop}%
\bibitem [{\citenamefont {Marukame}\ \emph {et~al.}(2006)\citenamefont
  {Marukame}, \citenamefont {Ishikawa}, \citenamefont {Hakamata},\ and\
  \citenamefont {Matsuda}}]{Marukame2006}%
  \BibitemOpen
  \bibfield  {author} {\bibinfo {author} {\bibfnamefont {T.}~\bibnamefont
  {Marukame}}, \bibinfo {author} {\bibfnamefont {T.}~\bibnamefont {Ishikawa}},
  \bibinfo {author} {\bibfnamefont {S.}~\bibnamefont {Hakamata}}, \ and\
  \bibinfo {author} {\bibfnamefont {K.-i.}\ \bibnamefont {Matsuda}},\ }\href
  {\doibase 10.1109/TMAG.2006.878859} {\bibfield  {journal} {\bibinfo
  {journal} {IEEE Trans. Magn.}\ }\textbf {\bibinfo {volume} {42}},\ \bibinfo
  {pages} {2652} (\bibinfo {year} {2006})}\BibitemShut {NoStop}%
\bibitem [{\citenamefont {Mavropoulos}\ \emph {et~al.}(2000)\citenamefont
  {Mavropoulos}, \citenamefont {Papanikolaou},\ and\ \citenamefont
  {Dederichs}}]{Mavropoulos2000a}%
  \BibitemOpen
  \bibfield  {author} {\bibinfo {author} {\bibfnamefont {P.}~\bibnamefont
  {Mavropoulos}}, \bibinfo {author} {\bibfnamefont {N.}~\bibnamefont
  {Papanikolaou}}, \ and\ \bibinfo {author} {\bibfnamefont {P.~H.}\
  \bibnamefont {Dederichs}},\ }\href {\doibase 10.1103/PhysRevLett.85.1088}
  {\bibfield  {journal} {\bibinfo  {journal} {Phys. Rev. Lett.}\ }\textbf
  {\bibinfo {volume} {85}},\ \bibinfo {pages} {1088} (\bibinfo {year}
  {2000})}\BibitemShut {NoStop}%
\bibitem [{\citenamefont {Miller}\ \emph {et~al.}(2006)\citenamefont {Miller},
  \citenamefont {Li}, \citenamefont {Schuller}, \citenamefont {Dave},
  \citenamefont {Slaughter},\ and\ \citenamefont {{\AA}kerman}}]{Miller2006}%
  \BibitemOpen
  \bibfield  {author} {\bibinfo {author} {\bibfnamefont {C.~W.}\ \bibnamefont
  {Miller}}, \bibinfo {author} {\bibfnamefont {Z.~P.}\ \bibnamefont {Li}},
  \bibinfo {author} {\bibfnamefont {I.~K.}\ \bibnamefont {Schuller}}, \bibinfo
  {author} {\bibfnamefont {R.~W.}\ \bibnamefont {Dave}}, \bibinfo {author}
  {\bibfnamefont {J.~M.}\ \bibnamefont {Slaughter}}, \ and\ \bibinfo {author}
  {\bibfnamefont {J.}~\bibnamefont {{\AA}kerman}},\ }\href {\doibase
  10.1103/PhysRevB.74.212404} {\bibfield  {journal} {\bibinfo  {journal} {Phys.
  Rev. B}\ }\textbf {\bibinfo {volume} {74}},\ \bibinfo {pages} {212404}
  (\bibinfo {year} {2006})}\BibitemShut {NoStop}%
\bibitem [{\citenamefont {Brinkman}\ \emph {et~al.}(1970)\citenamefont
  {Brinkman}, \citenamefont {Dynes},\ and\ \citenamefont
  {Rowell}}]{Brinkman1970}%
  \BibitemOpen
  \bibfield  {author} {\bibinfo {author} {\bibfnamefont {W.~F.}\ \bibnamefont
  {Brinkman}}, \bibinfo {author} {\bibfnamefont {R.~C.}\ \bibnamefont {Dynes}},
  \ and\ \bibinfo {author} {\bibfnamefont {J.~M.}\ \bibnamefont {Rowell}},\
  }\href {\doibase 10.1063/1.1659141} {\bibfield  {journal} {\bibinfo
  {journal} {J. Appl. Phys.}\ }\textbf {\bibinfo {volume} {41}},\ \bibinfo
  {pages} {1915} (\bibinfo {year} {1970})}\BibitemShut {NoStop}%
\bibitem [{\citenamefont {Miao}\ \emph {et~al.}(2009)\citenamefont {Miao},
  \citenamefont {Xiao},\ and\ \citenamefont {Gupta}}]{Miao2009}%
  \BibitemOpen
  \bibfield  {author} {\bibinfo {author} {\bibfnamefont {G.~X.}\ \bibnamefont
  {Miao}}, \bibinfo {author} {\bibfnamefont {G.}~\bibnamefont {Xiao}}, \ and\
  \bibinfo {author} {\bibfnamefont {A.}~\bibnamefont {Gupta}},\ }\href
  {\doibase 10.1209/0295-5075/87/47006} {\bibfield  {journal} {\bibinfo
  {journal} {Europhys. Lett.}\ }\textbf {\bibinfo {volume} {87}},\ \bibinfo
  {pages} {47006} (\bibinfo {year} {2009})}\BibitemShut {NoStop}%
\bibitem [{\citenamefont {Oliver}\ and\ \citenamefont
  {Nowak}(2004)}]{Oliver2004}%
  \BibitemOpen
  \bibfield  {author} {\bibinfo {author} {\bibfnamefont {B.}~\bibnamefont
  {Oliver}}\ and\ \bibinfo {author} {\bibfnamefont {J.}~\bibnamefont {Nowak}},\
  }\href {\doibase 10.1063/1.1631074} {\bibfield  {journal} {\bibinfo
  {journal} {J. Appl. Phys.}\ }\textbf {\bibinfo {volume} {95}},\ \bibinfo
  {pages} {546} (\bibinfo {year} {2004})}\BibitemShut {NoStop}%
\bibitem [{\citenamefont {Moodera}\ \emph {et~al.}(1998)\citenamefont
  {Moodera}, \citenamefont {Nowak},\ and\ \citenamefont
  {Veerdonk}}]{Moodera1998}%
  \BibitemOpen
  \bibfield  {author} {\bibinfo {author} {\bibfnamefont {J.~S.}\ \bibnamefont
  {Moodera}}, \bibinfo {author} {\bibfnamefont {J.}~\bibnamefont {Nowak}}, \
  and\ \bibinfo {author} {\bibfnamefont {R.~J. M. V.~D.}\ \bibnamefont
  {Veerdonk}},\ }\href {\doibase 10.1103/PhysRevLett.80.2941} {\bibfield
  {journal} {\bibinfo  {journal} {Phys. Rev. Lett.}\ }\textbf {\bibinfo
  {volume} {80}},\ \bibinfo {pages} {2941} (\bibinfo {year}
  {1998})}\BibitemShut {NoStop}%
\bibitem [{\citenamefont {Yoo}\ \emph {et~al.}(2010)\citenamefont {Yoo},
  \citenamefont {Jang}, \citenamefont {Prigodin}, \citenamefont {Kao},
  \citenamefont {Eom},\ and\ \citenamefont {Epstein}}]{Yoo2010}%
  \BibitemOpen
  \bibfield  {author} {\bibinfo {author} {\bibfnamefont {J.~W.}\ \bibnamefont
  {Yoo}}, \bibinfo {author} {\bibfnamefont {H.~W.}\ \bibnamefont {Jang}},
  \bibinfo {author} {\bibfnamefont {V.~N.}\ \bibnamefont {Prigodin}}, \bibinfo
  {author} {\bibfnamefont {C.}~\bibnamefont {Kao}}, \bibinfo {author}
  {\bibfnamefont {C.~B.}\ \bibnamefont {Eom}}, \ and\ \bibinfo {author}
  {\bibfnamefont {A.~J.}\ \bibnamefont {Epstein}},\ }\href {\doibase
  10.1016/j.synthmet.2009.11.019} {\bibfield  {journal} {\bibinfo  {journal}
  {Synth. Met.}\ }\textbf {\bibinfo {volume} {160}},\ \bibinfo {pages} {216}
  (\bibinfo {year} {2010})}\BibitemShut {NoStop}%
\bibitem [{\citenamefont {Appelbaom}(1967)}]{Appelbaom1960}%
  \BibitemOpen
  \bibfield  {author} {\bibinfo {author} {\bibfnamefont {J.~A.}\ \bibnamefont
  {Appelbaom}},\ }\href {\doibase 10.1103/PhysRev.154.633} {\bibfield
  {journal} {\bibinfo  {journal} {Phys. Rev.}\ }\textbf {\bibinfo {volume}
  {154}},\ \bibinfo {pages} {633} (\bibinfo {year} {1967})}\BibitemShut
  {NoStop}%
\bibitem [{\citenamefont {Ramsteiner}\ \emph {et~al.}(2008)\citenamefont
  {Ramsteiner}, \citenamefont {Brandt}, \citenamefont {Flissikowski},
  \citenamefont {Grahn}, \citenamefont {Hashimoto}, \citenamefont {Herfort},\
  and\ \citenamefont {Kostial}}]{Ramsteiner2008}%
  \BibitemOpen
  \bibfield  {author} {\bibinfo {author} {\bibfnamefont {M.}~\bibnamefont
  {Ramsteiner}}, \bibinfo {author} {\bibfnamefont {O.}~\bibnamefont {Brandt}},
  \bibinfo {author} {\bibfnamefont {T.}~\bibnamefont {Flissikowski}}, \bibinfo
  {author} {\bibfnamefont {H.~T.}\ \bibnamefont {Grahn}}, \bibinfo {author}
  {\bibfnamefont {M.}~\bibnamefont {Hashimoto}}, \bibinfo {author}
  {\bibfnamefont {J.}~\bibnamefont {Herfort}}, \ and\ \bibinfo {author}
  {\bibfnamefont {H.}~\bibnamefont {Kostial}},\ }\href {\doibase
  10.1103/PhysRevB.78.121303} {\bibfield  {journal} {\bibinfo  {journal} {Phys.
  Rev. B}\ }\textbf {\bibinfo {volume} {78}},\ \bibinfo {pages} {121303(R)}
  (\bibinfo {year} {2008})}\BibitemShut {NoStop}%
\bibitem [{\citenamefont {Lou}\ \emph {et~al.}(2007)\citenamefont {Lou},
  \citenamefont {Adelmann}, \citenamefont {Crooker}, \citenamefont {Garlid},
  \citenamefont {Zhang}, \citenamefont {Reddy}, \citenamefont {Flexner},
  \citenamefont {Palmstr{\o}m},\ and\ \citenamefont {Crowell}}]{Lou2007a}%
  \BibitemOpen
  \bibfield  {author} {\bibinfo {author} {\bibfnamefont {X.}~\bibnamefont
  {Lou}}, \bibinfo {author} {\bibfnamefont {C.}~\bibnamefont {Adelmann}},
  \bibinfo {author} {\bibfnamefont {S.~A.}\ \bibnamefont {Crooker}}, \bibinfo
  {author} {\bibfnamefont {E.~S.}\ \bibnamefont {Garlid}}, \bibinfo {author}
  {\bibfnamefont {J.}~\bibnamefont {Zhang}}, \bibinfo {author} {\bibfnamefont
  {K.~S.~M.}\ \bibnamefont {Reddy}}, \bibinfo {author} {\bibfnamefont {S.~D.}\
  \bibnamefont {Flexner}}, \bibinfo {author} {\bibfnamefont {C.~J.}\
  \bibnamefont {Palmstr{\o}m}}, \ and\ \bibinfo {author} {\bibfnamefont
  {P.~A.}\ \bibnamefont {Crowell}},\ }\href {\doibase 10.1038/nphys543}
  {\bibfield  {journal} {\bibinfo  {journal} {Nat. Phys.}\ }\textbf {\bibinfo
  {volume} {3}},\ \bibinfo {pages} {197} (\bibinfo {year} {2007})}\BibitemShut
  {NoStop}%
\bibitem [{\citenamefont {Uemura}\ \emph {et~al.}(2011)\citenamefont {Uemura},
  \citenamefont {Akiho}, \citenamefont {Harada}, \citenamefont {Matsuda},\ and\
  \citenamefont {Yamamoto}}]{Uemura2011a}%
  \BibitemOpen
  \bibfield  {author} {\bibinfo {author} {\bibfnamefont {T.}~\bibnamefont
  {Uemura}}, \bibinfo {author} {\bibfnamefont {T.}~\bibnamefont {Akiho}},
  \bibinfo {author} {\bibfnamefont {M.}~\bibnamefont {Harada}}, \bibinfo
  {author} {\bibfnamefont {K.~I.}\ \bibnamefont {Matsuda}}, \ and\ \bibinfo
  {author} {\bibfnamefont {M.}~\bibnamefont {Yamamoto}},\ }\href {\doibase
  10.1063/1.3630032} {\bibfield  {journal} {\bibinfo  {journal} {Appl. Phys.
  Lett.}\ }\textbf {\bibinfo {volume} {99}},\ \bibinfo {pages} {082108}
  (\bibinfo {year} {2011})}\BibitemShut {NoStop}%
\bibitem [{\citenamefont {Ciorga}\ \emph {et~al.}(2009)\citenamefont {Ciorga},
  \citenamefont {Einwanger}, \citenamefont {Wurstbauer}, \citenamefont {Schuh},
  \citenamefont {Wegscheider},\ and\ \citenamefont {Weiss}}]{Ciorga2009}%
  \BibitemOpen
  \bibfield  {author} {\bibinfo {author} {\bibfnamefont {M.}~\bibnamefont
  {Ciorga}}, \bibinfo {author} {\bibfnamefont {A.}~\bibnamefont {Einwanger}},
  \bibinfo {author} {\bibfnamefont {U.}~\bibnamefont {Wurstbauer}}, \bibinfo
  {author} {\bibfnamefont {D.}~\bibnamefont {Schuh}}, \bibinfo {author}
  {\bibfnamefont {W.}~\bibnamefont {Wegscheider}}, \ and\ \bibinfo {author}
  {\bibfnamefont {D.}~\bibnamefont {Weiss}},\ }\href {\doibase
  10.1103/PhysRevB.79.165321} {\bibfield  {journal} {\bibinfo  {journal} {Phys.
  Rev. B}\ }\textbf {\bibinfo {volume} {79}},\ \bibinfo {pages} {165321}
  (\bibinfo {year} {2009})}\BibitemShut {NoStop}%
\bibitem [{\citenamefont {Salis}\ \emph {et~al.}(2009)\citenamefont {Salis},
  \citenamefont {Fuhrer},\ and\ \citenamefont {Alvarado}}]{Salis2009}%
  \BibitemOpen
  \bibfield  {author} {\bibinfo {author} {\bibfnamefont {G.}~\bibnamefont
  {Salis}}, \bibinfo {author} {\bibfnamefont {A.}~\bibnamefont {Fuhrer}}, \
  and\ \bibinfo {author} {\bibfnamefont {S.}~\bibnamefont {Alvarado}},\ }\href
  {\doibase 10.1103/PhysRevB.80.115332} {\bibfield  {journal} {\bibinfo
  {journal} {Phys. Rev. B}\ }\textbf {\bibinfo {volume} {80}},\ \bibinfo
  {pages} {115332} (\bibinfo {year} {2009})}\BibitemShut {NoStop}%
\bibitem [{\citenamefont {Sasaki}\ \emph {et~al.}(2010)\citenamefont {Sasaki},
  \citenamefont {Oikawa}, \citenamefont {Suzuki}, \citenamefont {Shiraishi},
  \citenamefont {Suzuki},\ and\ \citenamefont {Noguchi}}]{Sasaki2010}%
  \BibitemOpen
  \bibfield  {author} {\bibinfo {author} {\bibfnamefont {T.}~\bibnamefont
  {Sasaki}}, \bibinfo {author} {\bibfnamefont {T.}~\bibnamefont {Oikawa}},
  \bibinfo {author} {\bibfnamefont {T.}~\bibnamefont {Suzuki}}, \bibinfo
  {author} {\bibfnamefont {M.}~\bibnamefont {Shiraishi}}, \bibinfo {author}
  {\bibfnamefont {Y.}~\bibnamefont {Suzuki}}, \ and\ \bibinfo {author}
  {\bibfnamefont {K.}~\bibnamefont {Noguchi}},\ }\href {\doibase
  10.1143/APEX.2.053003} {\bibfield  {journal} {\bibinfo  {journal} {IEEE
  Trans. Magn.}\ }\textbf {\bibinfo {volume} {46}},\ \bibinfo {pages} {1436}
  (\bibinfo {year} {2010})}\BibitemShut {NoStop}%
\bibitem [{\citenamefont {Ji}\ \emph {et~al.}(2007)\citenamefont {Ji},
  \citenamefont {Hoffmann}, \citenamefont {Jiang}, \citenamefont {Pearson},\
  and\ \citenamefont {Bader}}]{Ji2007}%
  \BibitemOpen
  \bibfield  {author} {\bibinfo {author} {\bibfnamefont {Y.}~\bibnamefont
  {Ji}}, \bibinfo {author} {\bibfnamefont {A.}~\bibnamefont {Hoffmann}},
  \bibinfo {author} {\bibfnamefont {J.~S.}\ \bibnamefont {Jiang}}, \bibinfo
  {author} {\bibfnamefont {J.~E.}\ \bibnamefont {Pearson}}, \ and\ \bibinfo
  {author} {\bibfnamefont {S.~D.}\ \bibnamefont {Bader}},\ }\href {\doibase
  10.1088/0022-3727/40/5/S13} {\bibfield  {journal} {\bibinfo  {journal} {J.
  Phys. D. Appl. Phys.}\ }\textbf {\bibinfo {volume} {40}},\ \bibinfo {pages}
  {1280} (\bibinfo {year} {2007})}\BibitemShut {NoStop}%
\bibitem [{\citenamefont {Fabian}\ \emph {et~al.}(2007)\citenamefont {Fabian},
  \citenamefont {Matos-Abiague}, \citenamefont {Ertler}, \citenamefont
  {Stano},\ and\ \citenamefont {{\v{Z}}uti{\'{c}}}}]{Fabian2007}%
  \BibitemOpen
  \bibfield  {author} {\bibinfo {author} {\bibfnamefont {J.}~\bibnamefont
  {Fabian}}, \bibinfo {author} {\bibfnamefont {A.}~\bibnamefont
  {Matos-Abiague}}, \bibinfo {author} {\bibfnamefont {C.}~\bibnamefont
  {Ertler}}, \bibinfo {author} {\bibfnamefont {P.}~\bibnamefont {Stano}}, \
  and\ \bibinfo {author} {\bibfnamefont {I.}~\bibnamefont
  {{\v{Z}}uti{\'{c}}}},\ }\href {\doibase 10.2478/v10155-010-0086-8} {\bibfield
   {journal} {\bibinfo  {journal} {Acta Phys. Slovaca. Rev. Tutorials}\
  }\textbf {\bibinfo {volume} {57}},\ \bibinfo {pages} {50} (\bibinfo {year}
  {2007})}\BibitemShut {NoStop}%
\bibitem [{\citenamefont {Johnson}\ and\ \citenamefont
  {Silsbee}(1988)}]{Johnson1988}%
  \BibitemOpen
  \bibfield  {author} {\bibinfo {author} {\bibfnamefont {M.}~\bibnamefont
  {Johnson}}\ and\ \bibinfo {author} {\bibfnamefont {R.~H.}\ \bibnamefont
  {Silsbee}},\ }\href {\doibase 10.1103/PhysRevB.37.5312} {\bibfield  {journal}
  {\bibinfo  {journal} {Phys. Rev. B}\ }\textbf {\bibinfo {volume} {37}},\
  \bibinfo {pages} {5312} (\bibinfo {year} {1988})}\BibitemShut {NoStop}%
\bibitem [{\citenamefont {Fabian}\ and\ \citenamefont {Zutic}(2009)}]{Fabian}%
  \BibitemOpen
  \bibfield  {author} {\bibinfo {author} {\bibfnamefont {J.}~\bibnamefont
  {Fabian}}\ and\ \bibinfo {author} {\bibfnamefont {I.}~\bibnamefont {Zutic}},\
  }\href@noop {} {\emph {\bibinfo {title} {Stand. Model spin Inject.}}},\
  \bibinfo {type} {Tech. Rep.}\ (\bibinfo  {institution} {Institute for
  Theoretical Physics University of Regensburg D-93040 Regensburg, Germany},\
  \bibinfo {year} {2009})\ \Eprint {http://arxiv.org/abs/arXiv:0903.2500v1}
  {arXiv:arXiv:0903.2500v1} \BibitemShut {NoStop}%
\bibitem [{\citenamefont {Chan}\ \emph {et~al.}(2009)\citenamefont {Chan},
  \citenamefont {Hu}, \citenamefont {Zhang}, \citenamefont {Kondo},
  \citenamefont {Palmstr{\o}m},\ and\ \citenamefont {Crowell}}]{Chan2009}%
  \BibitemOpen
  \bibfield  {author} {\bibinfo {author} {\bibfnamefont {M.~K.}\ \bibnamefont
  {Chan}}, \bibinfo {author} {\bibfnamefont {Q.~O.}\ \bibnamefont {Hu}},
  \bibinfo {author} {\bibfnamefont {J.}~\bibnamefont {Zhang}}, \bibinfo
  {author} {\bibfnamefont {T.}~\bibnamefont {Kondo}}, \bibinfo {author}
  {\bibfnamefont {C.~J.}\ \bibnamefont {Palmstr{\o}m}}, \ and\ \bibinfo
  {author} {\bibfnamefont {P.~A.}\ \bibnamefont {Crowell}},\ }\href {\doibase
  10.1103/PhysRevB.80.161206} {\bibfield  {journal} {\bibinfo  {journal} {Phys.
  Rev. B}\ }\textbf {\bibinfo {volume} {80}},\ \bibinfo {pages} {161206(R)}
  (\bibinfo {year} {2009})}\BibitemShut {NoStop}%
\bibitem [{\citenamefont {Shiogai}\ \emph {et~al.}(2012)\citenamefont
  {Shiogai}, \citenamefont {Ciorga}, \citenamefont {Utz}, \citenamefont
  {Schuh}, \citenamefont {Arakawa}, \citenamefont {Kohda}, \citenamefont
  {Kobayashi}, \citenamefont {Ono}, \citenamefont {Wegscheider}, \citenamefont
  {Weiss},\ and\ \citenamefont {Nitta}}]{Shiogai2012a}%
  \BibitemOpen
  \bibfield  {author} {\bibinfo {author} {\bibfnamefont {J.}~\bibnamefont
  {Shiogai}}, \bibinfo {author} {\bibfnamefont {M.}~\bibnamefont {Ciorga}},
  \bibinfo {author} {\bibfnamefont {M.}~\bibnamefont {Utz}}, \bibinfo {author}
  {\bibfnamefont {D.}~\bibnamefont {Schuh}}, \bibinfo {author} {\bibfnamefont
  {T.}~\bibnamefont {Arakawa}}, \bibinfo {author} {\bibfnamefont
  {M.}~\bibnamefont {Kohda}}, \bibinfo {author} {\bibfnamefont
  {K.}~\bibnamefont {Kobayashi}}, \bibinfo {author} {\bibfnamefont
  {T.}~\bibnamefont {Ono}}, \bibinfo {author} {\bibfnamefont {W.}~\bibnamefont
  {Wegscheider}}, \bibinfo {author} {\bibfnamefont {D.}~\bibnamefont {Weiss}},
  \ and\ \bibinfo {author} {\bibfnamefont {J.}~\bibnamefont {Nitta}},\ }\href
  {\doibase 10.1063/1.4767339} {\bibfield  {journal} {\bibinfo  {journal}
  {Appl. Phys. Lett.}\ }\textbf {\bibinfo {volume} {101}},\ \bibinfo {pages}
  {212402} (\bibinfo {year} {2012})}\BibitemShut {NoStop}%
\bibitem [{\citenamefont {Salis}\ \emph {et~al.}(2010)\citenamefont {Salis},
  \citenamefont {Fuhrer}, \citenamefont {Schlittler}, \citenamefont {Gross},\
  and\ \citenamefont {Alvarado}}]{Salis2010}%
  \BibitemOpen
  \bibfield  {author} {\bibinfo {author} {\bibfnamefont {G.}~\bibnamefont
  {Salis}}, \bibinfo {author} {\bibfnamefont {A.}~\bibnamefont {Fuhrer}},
  \bibinfo {author} {\bibfnamefont {R.~R.}\ \bibnamefont {Schlittler}},
  \bibinfo {author} {\bibfnamefont {L.}~\bibnamefont {Gross}}, \ and\ \bibinfo
  {author} {\bibfnamefont {S.~F.}\ \bibnamefont {Alvarado}},\ }\href {\doibase
  10.1103/PhysRevB.81.205323} {\bibfield  {journal} {\bibinfo  {journal} {Phys.
  Rev. B}\ }\textbf {\bibinfo {volume} {81}},\ \bibinfo {pages} {205323}
  (\bibinfo {year} {2010})}\BibitemShut {NoStop}%
\bibitem [{\citenamefont {Hashimoto}\ \emph {et~al.}(2006)\citenamefont
  {Hashimoto}, \citenamefont {Herfort}, \citenamefont {Trampert}, \citenamefont
  {Sch{\"{o}}nherr},\ and\ \citenamefont {Ploog}}]{Hashimoto2006a}%
  \BibitemOpen
  \bibfield  {author} {\bibinfo {author} {\bibfnamefont {M.}~\bibnamefont
  {Hashimoto}}, \bibinfo {author} {\bibfnamefont {J.}~\bibnamefont {Herfort}},
  \bibinfo {author} {\bibfnamefont {A.}~\bibnamefont {Trampert}}, \bibinfo
  {author} {\bibfnamefont {H.-P.}\ \bibnamefont {Sch{\"{o}}nherr}}, \ and\
  \bibinfo {author} {\bibfnamefont {K.~H.}\ \bibnamefont {Ploog}},\ }\href
  {\doibase 10.1116/1.2218863} {\bibfield  {journal} {\bibinfo  {journal} {J.
  Vac. Sci. Technol. B}\ }\textbf {\bibinfo {volume} {24}},\ \bibinfo {pages}
  {2004} (\bibinfo {year} {2006})}\BibitemShut {NoStop}%
\bibitem [{\citenamefont {Fluitman}(1973)}]{Fluitman1973a}%
  \BibitemOpen
  \bibfield  {author} {\bibinfo {author} {\bibfnamefont {J.~H.~J.}\
  \bibnamefont {Fluitman}},\ }\href {\doibase 10.1016/0040-6090(73)90080-1}
  {\bibfield  {journal} {\bibinfo  {journal} {Thin Sol. Films}\ }\textbf
  {\bibinfo {volume} {16}},\ \bibinfo {pages} {269} (\bibinfo {year}
  {1973})}\BibitemShut {NoStop}%
\bibitem [{\citenamefont {Kryder}\ \emph {et~al.}(1980)\citenamefont {Kryder},
  \citenamefont {Ahn}, \citenamefont {Mazzeo}, \citenamefont {Schwarzl},\ and\
  \citenamefont {Kane}}]{Kryder1980a}%
  \BibitemOpen
  \bibfield  {author} {\bibinfo {author} {\bibfnamefont {M.~H.}\ \bibnamefont
  {Kryder}}, \bibinfo {author} {\bibfnamefont {K.~Y.}\ \bibnamefont {Ahn}},
  \bibinfo {author} {\bibfnamefont {N.~J.}\ \bibnamefont {Mazzeo}}, \bibinfo
  {author} {\bibfnamefont {S.}~\bibnamefont {Schwarzl}}, \ and\ \bibinfo
  {author} {\bibfnamefont {S.~M.}\ \bibnamefont {Kane}},\ }\href {\doibase
  10.1109/TMAG.1980.1060555} {\bibfield  {journal} {\bibinfo  {journal} {IEEE
  Trans. Magn.}\ }\textbf {\bibinfo {volume} {MAG-16}},\ \bibinfo {pages} {99}
  (\bibinfo {year} {1980})}\BibitemShut {NoStop}%
\bibitem [{\citenamefont {Lee}\ \emph {et~al.}(1999)\citenamefont {Lee},
  \citenamefont {Xu}, \citenamefont {Vaz}, \citenamefont {Hirohata},
  \citenamefont {Leung}, \citenamefont {Yao}, \citenamefont {Choi},
  \citenamefont {Bland}, \citenamefont {Rousseaux}, \citenamefont {Cambril},\
  and\ \citenamefont {Launois}}]{Lee1999}%
  \BibitemOpen
  \bibfield  {author} {\bibinfo {author} {\bibfnamefont {W.}~\bibnamefont
  {Lee}}, \bibinfo {author} {\bibfnamefont {Y.}~\bibnamefont {Xu}}, \bibinfo
  {author} {\bibfnamefont {C.}~\bibnamefont {Vaz}}, \bibinfo {author}
  {\bibfnamefont {A.}~\bibnamefont {Hirohata}}, \bibinfo {author}
  {\bibfnamefont {H.}~\bibnamefont {Leung}}, \bibinfo {author} {\bibfnamefont
  {C.}~\bibnamefont {Yao}}, \bibinfo {author} {\bibfnamefont {B.-C.}\
  \bibnamefont {Choi}}, \bibinfo {author} {\bibfnamefont {J.}~\bibnamefont
  {Bland}}, \bibinfo {author} {\bibfnamefont {F.}~\bibnamefont {Rousseaux}},
  \bibinfo {author} {\bibfnamefont {E.}~\bibnamefont {Cambril}}, \ and\
  \bibinfo {author} {\bibfnamefont {H.}~\bibnamefont {Launois}},\ }\href
  {\doibase 10.1109/20.800696} {\bibfield  {journal} {\bibinfo  {journal} {IEEE
  Trans. Magn.}\ }\textbf {\bibinfo {volume} {35}},\ \bibinfo {pages} {3883}
  (\bibinfo {year} {1999})}\BibitemShut {NoStop}%
\bibitem [{\citenamefont {Saikin}\ \emph {et~al.}(2006)\citenamefont {Saikin},
  \citenamefont {Shen},\ and\ \citenamefont {Cheng}}]{Saikin2006a}%
  \BibitemOpen
  \bibfield  {author} {\bibinfo {author} {\bibfnamefont {S.}~\bibnamefont
  {Saikin}}, \bibinfo {author} {\bibfnamefont {M.}~\bibnamefont {Shen}}, \ and\
  \bibinfo {author} {\bibfnamefont {M.~C.}\ \bibnamefont {Cheng}},\ }\href
  {\doibase 10.1088/0953-8984/18/5/005} {\bibfield  {journal} {\bibinfo
  {journal} {J. Phys. Condens. Matter}\ }\textbf {\bibinfo {volume} {18}},\
  \bibinfo {pages} {1535} (\bibinfo {year} {2006})}\BibitemShut {NoStop}%
\bibitem [{\citenamefont {Song}\ and\ \citenamefont {Dery}(2010)}]{Song2010a}%
  \BibitemOpen
  \bibfield  {author} {\bibinfo {author} {\bibfnamefont {Y.}~\bibnamefont
  {Song}}\ and\ \bibinfo {author} {\bibfnamefont {H.}~\bibnamefont {Dery}},\
  }\href {\doibase 10.1103/PhysRevB.81.045321} {\bibfield  {journal} {\bibinfo
  {journal} {Phys. Rev. B}\ }\textbf {\bibinfo {volume} {81}},\ \bibinfo
  {pages} {045321} (\bibinfo {year} {2010})}\BibitemShut {NoStop}%
\bibitem [{\citenamefont {Salis}\ \emph {et~al.}(2011)\citenamefont {Salis},
  \citenamefont {Alvarado},\ and\ \citenamefont {Fuhrer}}]{Salis2011a}%
  \BibitemOpen
  \bibfield  {author} {\bibinfo {author} {\bibfnamefont {G.}~\bibnamefont
  {Salis}}, \bibinfo {author} {\bibfnamefont {S.~F.}\ \bibnamefont {Alvarado}},
  \ and\ \bibinfo {author} {\bibfnamefont {A.}~\bibnamefont {Fuhrer}},\ }\href
  {\doibase 10.1103/PhysRevB.84.041307} {\bibfield  {journal} {\bibinfo
  {journal} {Phys. Rev. B}\ }\textbf {\bibinfo {volume} {84}},\ \bibinfo
  {pages} {041307} (\bibinfo {year} {2011})}\BibitemShut {NoStop}%
\bibitem [{\citenamefont {Valenzuela}\ \emph {et~al.}(2005)\citenamefont
  {Valenzuela}, \citenamefont {Monsma}, \citenamefont {Marcus}, \citenamefont
  {Narayanamurti},\ and\ \citenamefont {Tinkham}}]{Valenzuela2005a}%
  \BibitemOpen
  \bibfield  {author} {\bibinfo {author} {\bibfnamefont {S.~O.}\ \bibnamefont
  {Valenzuela}}, \bibinfo {author} {\bibfnamefont {D.~J.}\ \bibnamefont
  {Monsma}}, \bibinfo {author} {\bibfnamefont {C.~M.}\ \bibnamefont {Marcus}},
  \bibinfo {author} {\bibfnamefont {V.}~\bibnamefont {Narayanamurti}}, \ and\
  \bibinfo {author} {\bibfnamefont {M.}~\bibnamefont {Tinkham}},\ }\href
  {\doibase 10.1103/PhysRevLett.94.196601} {\bibfield  {journal} {\bibinfo
  {journal} {Phys. Rev. Lett.}\ }\textbf {\bibinfo {volume} {94}},\ \bibinfo
  {pages} {196601} (\bibinfo {year} {2005})}\BibitemShut {NoStop}%
\bibitem [{\citenamefont {{van 't Erve}}\ \emph {et~al.}(2012)\citenamefont
  {{van 't Erve}}, \citenamefont {Friedman}, \citenamefont {Cobas},
  \citenamefont {Li}, \citenamefont {Robinson},\ and\ \citenamefont
  {Jonker}}]{VantErve2012a}%
  \BibitemOpen
  \bibfield  {author} {\bibinfo {author} {\bibfnamefont {O.~M.~J.}\
  \bibnamefont {{van 't Erve}}}, \bibinfo {author} {\bibfnamefont {A.~L.}\
  \bibnamefont {Friedman}}, \bibinfo {author} {\bibfnamefont {E.}~\bibnamefont
  {Cobas}}, \bibinfo {author} {\bibfnamefont {C.~H.}\ \bibnamefont {Li}},
  \bibinfo {author} {\bibfnamefont {J.~T.}\ \bibnamefont {Robinson}}, \ and\
  \bibinfo {author} {\bibfnamefont {B.~T.}\ \bibnamefont {Jonker}},\ }\href
  {\doibase 10.1038/nnano.2012.161} {\bibfield  {journal} {\bibinfo  {journal}
  {Nat. Nanotechnol.}\ }\textbf {\bibinfo {volume} {7}},\ \bibinfo {pages}
  {737} (\bibinfo {year} {2012})}\BibitemShut {NoStop}%
\bibitem [{\citenamefont {Bruski}\ \emph {et~al.}(2014)\citenamefont {Bruski},
  \citenamefont {Erwin}, \citenamefont {Herfort}, \citenamefont {Tahraoui},\
  and\ \citenamefont {Ramsteiner}}]{Bruski2014}%
  \BibitemOpen
  \bibfield  {author} {\bibinfo {author} {\bibfnamefont {P.}~\bibnamefont
  {Bruski}}, \bibinfo {author} {\bibfnamefont {S.~C.}\ \bibnamefont {Erwin}},
  \bibinfo {author} {\bibfnamefont {J.}~\bibnamefont {Herfort}}, \bibinfo
  {author} {\bibfnamefont {A.}~\bibnamefont {Tahraoui}}, \ and\ \bibinfo
  {author} {\bibfnamefont {M.}~\bibnamefont {Ramsteiner}},\ }\href {\doibase
  10.1103/PhysRevB.90.245150} {\bibfield  {journal} {\bibinfo  {journal} {Phys.
  Rev. B}\ }\textbf {\bibinfo {volume} {90}},\ \bibinfo {pages} {245150}
  (\bibinfo {year} {2014})}\BibitemShut {NoStop}%
\bibitem [{\citenamefont {Bruski}\ \emph {et~al.}(2011)\citenamefont {Bruski},
  \citenamefont {Erwin}, \citenamefont {Ramsteiner}, \citenamefont {Brandt},
  \citenamefont {Friedland}, \citenamefont {Farshchi}, \citenamefont
  {Herfort},\ and\ \citenamefont {Riechert}}]{Bruski2011a}%
  \BibitemOpen
  \bibfield  {author} {\bibinfo {author} {\bibfnamefont {P.}~\bibnamefont
  {Bruski}}, \bibinfo {author} {\bibfnamefont {S.}~\bibnamefont {Erwin}},
  \bibinfo {author} {\bibfnamefont {M.}~\bibnamefont {Ramsteiner}}, \bibinfo
  {author} {\bibfnamefont {O.}~\bibnamefont {Brandt}}, \bibinfo {author}
  {\bibfnamefont {K.-J.}\ \bibnamefont {Friedland}}, \bibinfo {author}
  {\bibfnamefont {R.}~\bibnamefont {Farshchi}}, \bibinfo {author}
  {\bibfnamefont {J.}~\bibnamefont {Herfort}}, \ and\ \bibinfo {author}
  {\bibfnamefont {H.}~\bibnamefont {Riechert}},\ }\href {\doibase
  10.1103/PhysRevB.83.140409} {\bibfield  {journal} {\bibinfo  {journal} {Phys.
  Rev. B}\ }\textbf {\bibinfo {volume} {83}},\ \bibinfo {pages} {140409(R)}
  (\bibinfo {year} {2011})}\BibitemShut {NoStop}%
\end{thebibliography}
\end{document}